%% file: RIS-RQSM-12.24-2022_-_w.o_notes.tex
\batchmode
\makeatletter
\def\input@path{{"F:/My Drive/UCD/Writing/RIS-RQSM/ArXiv/"}}
\makeatother
\documentclass[12pt, draftclsnofoot, onecolumn]{IEEEtran}
\usepackage[T1]{fontenc}
\usepackage[latin9]{inputenc}
\usepackage{array}
\usepackage{amsmath}
\usepackage{amsthm}
\usepackage{amssymb}
\usepackage{graphicx}
\usepackage[pdftex,unicode=true,
 bookmarks=true,bookmarksnumbered=true,bookmarksopen=true,bookmarksopenlevel=1,
 breaklinks=false,pdfborder={0 0 0},pdfborderstyle={},backref=false,colorlinks=false]
 {hyperref}
\hypersetup{pdftitle={Your Title},
 pdfauthor={Your Name},
 pdfpagelayout=OneColumn, pdfnewwindow=true, pdfstartview=XYZ, plainpages=false}

\makeatletter

\providecommand{\tabularnewline}{\\}

\theoremstyle{plain}
\newtheorem{thm}{\protect\theoremname}

\usepackage[caption=false,font=footnotesize]{subfig}
\usepackage{acronym}
\AtBeginDocument{\input{acro.tex}}

\makeatother

\providecommand{\theoremname}{Theorem}

\begin{document}
\title{RIS-Assisted Receive Quadrature Spatial Modulation with Low-Complexity\\Greedy
Detection}
\author{Mohamad H. Dinan,~\IEEEmembership{Member,~IEEE}, Marco Di Renzo,~\IEEEmembership{Fellow,~IEEE},\\and~Mark
F. Flanagan,~\IEEEmembership{Senior~Member,~IEEE}\thanks{This work was funded by the Irish Research Council (IRC) under the
Consolidator Laureate Award Programme (grant number IRCLA/2017/209).
The work of Marco Di Renzo was supported in part by the European Commission
through the H2020 ARIADNE project under grant agreement number 871464
and through the H2020 RISE-6G project under grant agreement number
101017011.}\thanks{Mohamad H. Dinan and Mark F. Flanagan are with the School of Electrical
and Electronic Engineering, University College Dublin, Belfield, Dublin
4, D04 V1W8 Ireland (email: \protect\href{mailto:mohamad.hejazidinan@ucdconnect.ie}{mohamad.hejazidinan@ucdconnect.ie};
\protect\href{mailto:mark.flanagan@ieee.org}{mark.flanagan@ieee.org}).}\thanks{Marco Di Renzo is with Universit\'e Paris-Saclay, CNRS, CentraleSup\'elec,
Laboratoire des Signaux et Syst\`emes, 3 Rue Joliot-Curie, 91192
Gif-sur-Yvette, France (email: \protect\href{mailto:marco.di-renzo@universite-paris-saclay.fr}{marco.di-renzo@universite-paris-saclay.fr}).}}
\maketitle
\begin{abstract}
In this paper, we propose a novel reconfigurable intelligent surface
(RIS)-assisted wireless communication scheme which uses the concept
of spatial modulation, namely RIS-assisted receive quadrature spatial
modulation (RIS-RQSM). In the proposed RIS-RQSM system, the information
bits are conveyed via both the indices of the \emph{two} selected
receive antennas \emph{and} the conventional in-phase/quadrature
(IQ) modulation. We propose a novel methodology to adjust the phase
shifts of the RIS elements in order to maximize the signal-to-noise
ratio (SNR) \emph{and} at the same time to \emph{construct} two separate
PAM symbols at the selected receive antennas, as the in-phase and
quadrature components of the desired IQ symbol. An energy-based greedy
detector (GD) is implemented at the receiver to efficiently detect
the received signal with minimal channel state information (CSI) via
the use of an appropriately designed one-tap pre-equalizer. We also
derive a closed-form upper bound on the average bit error probability
(ABEP) of the proposed RIS-RQSM system. Then, we formulate an optimization
problem to minimize the ABEP in order to improve the performance of
the system, which allows the GD to act as a near-optimal receiver.
Extensive numerical results are provided to demonstrate the error
rate performance of the system and to compare with that of a prominent
benchmark scheme. The results verify the remarkable superiority of
the proposed RIS-RQSM system over the benchmark scheme.
\end{abstract}


\section{Introduction\label{sec:Introduction}}

In the past few years, various wireless communication technologies
have emerged with an aim to support high demands for connectivity
and an immense increase in mobile data traffic. Among these, reconfigurable
intelligent surfaces (RISs), also known as intelligent reflecting
surfaces (IRSs), represents a key innovation that has drawn significant
attention from researchers in both academia and industry \cite{liu2022path}
and is foreseen to be a potential candidate for 6th generation (6G)
networks \cite{di2020smart,alghamdi2020intelligent}. An RIS is a
surface of electromagnetic meta-material consisting of a large number
of small, low-cost and energy-efficient reflecting elements that are
able to control the scattering and propagation in the channel by inducing
a pre-designed phase shift to the impinging wave. From this perspective,
RIS technology represents a revolutionary paradigm that can transform
the uncontrollable disruptive propagation environment into a \emph{smart radio environment}
\cite{di2020smart,di2019smart}, thus enhancing the received signal
quality \cite{basar2019transmission,basar2019wireless}.

On the other hand, spatial modulation (SM) \cite{mesleh2008spatial,di2011spatial}
and its variants such as generalized spatial modulation (GSM) \cite{younis2010generalized},
receive spatial modulation (RSM) \cite{yang2011preprocessing,stavridis2012precoding},
and quadrature spatial modulation (QSM) \cite{mesleh2014quadrature},
have been widely investigated in the last two decades as a promising
technology for beyond-5th-generation (B5G) networks. SM uses the \emph{indices}
of the transmit/receive antennas to convey the information bits. It
exploits the channel attributes to simplify the transceiver structure
in order to provide a more energy-efficient solution compared with
other conventional multiple-input multiple-output (MIMO) techniques
\cite{di2014spatial}.

The implicit advantages of both RIS and SM technology have motivated
researchers to combine these two advanced technologies to obtain a
reliable energy-efficient approach in order to achieve so-called green
or sustainable wireless communications. Specifically, in \cite{basar2020reconfigurable},
two fundamental RIS-based index modulation (IM) techniques were proposed,
i.e., RIS-space-shift keying (RIS-SSK) and RIS-spatial modulation
(RIS-SM). In both scenarios, the RIS-access point (RIS-AP) approach
was implemented, in which the RIS forms part of the transmitter, and
the index of the \emph{receive} antennas is used to convey the data
bits. The numerical results confirm a significant superiority of these
RIS-aided schemes compared to conventional MIMO schemes. Various principles
of RIS-based SM (also known as metasurface-based modulation) were
introduced in \cite{li2021single}. The authors of \cite{canbilen2022performance}
proposed an RIS-SSK system with multiple transmit antennas in which
the information bits map to the \emph{transmit} antenna index and
the single-antenna receiver receives the signal reflected from the
RIS. Various scenarios with ideal and non-ideal transceivers were
investigated and the error rate performance of each scenario was analyzed.
The results indicate that maximizing the signal-to-noise ratio (SNR)
at the receiver is not a good approach for the transmit RIS-SSK setting,
and in fact shows a relatively poor performance. In light of this,
in \cite{li2021space} the authors proposed an optimization algorithm
for the transmit RIS-SSK system to maximize the minimum Euclidean
distance among the received symbols. Using this approach, a performance
improvement is achieved at the expense of an increased computational
complexity. Moreover, in \cite{Luo2021spatial}, adopting a similar
approach, the authors proposed a joint optimization of the power allocation
matrix and the phase shifts of the RIS elements. An RIS-based SM system
with both the transmit and receive antenna index modulation was proposed
in \cite{ma2020large} to increase the spectral efficiency. However,
the results show that the error rate performance of the transmit SM
bits is extensively lower than that of the receive SM bits; this is
due to a reduction in the resulting channel-imprinted Euclidean distances.
RIS-aided receive quadrature reflecting modulation (RIS-RQRM) proposed
in \cite{yuan2021receive} is another interesting approach in which
QSM is applied within the receive antenna array. In this scenario,
the RIS is divided into two halves, and each half targets the real
or imaginary part of the signal at the two selected receive antennas
in order to double the throughput; however, the SNR at the receiver
is significantly reduced due the reduction in the number of RIS elements
per targeted antenna. Inspired by RIS-RQRM and \cite{ma2020large},
an IRS-assisted transceiver QSM (IRS-TQSM) scheme was proposed in
\cite{sanila2022tqsm} which applies QSM at both the transmitter and
the receiver. In \cite{zhang2022gssk,albinsaid2021multiple}, generalized
SSK (GSSK) and GSM approaches have been implemented in an RIS-assisted
wireless system. In both scenarios, the RIS is divided into multiple
parts to target multiple antennas at the receiver; hence, the throughput
can be increased at the expense of a decrease in the SNR at the target
antennas. The concept of SM has also been applied within the RIS entity
in \cite{lin2020reconfigurable,lin2021reconfigurable_journal,shu2022beamformin}
in order to transmit additional data bits. This is an exciting approach
to transmit the environmental data collected by the RIS; however,
experimental results show a very large degradation in the error rate
performance of the SM symbol, that is due to the similarity within
the possible (noise-free) received signals. In order to tackle the
problem of the SNR decrease due to grouping of the RIS elements, in
\cite{dinan2022ris} we proposed a new paradigm, namely RIS-assisted
receive quadrature space-shift keying (RIS-RQSSK) in order to simultaneously
target \emph{two} receive antennas. An optimization problem was defined
to maximize the SNR of the real part of the signal at one antenna
and, at the same time, of the imaginary part of the signal at the
second antenna. The spectral efficiency of this approach is increased
without any degradation in the SNR. However, the throughput of the
RIS-RQSSK system is limited and can only be increased by increasing
the number of receive antennas which is not a viable option in practice.

Against this background, in this paper we introduce a new RIS-assisted
quadrature scheme in which, in addition to mapping the information
bits independently to two indices of receive antennas, additional
bits are transmitted via conventional in-phase/quadrature (IQ) modulation.
The contributions of this paper are as follows:
\begin{itemize}
\item To improve the spectral efficiency of RIS-RQSSK while preserving its
excellent performance, we propose an RIS-assisted receive quadrature
spatial modulation (RIS-RQSM) system. In particular, all RIS elements
target two independently selected receive antennas to convey the information
bits. In this scenario, we introduce a novel idea to optimize the
phase shifts of the RIS elements in order to not only maximize the
SNR components associated to the real and imaginary parts of the signal
at the receive antennas, but also to help in constructing the in-phase
(I) and quadrature (Q) components of the symbol at the two separate
antennas. Specifically, the phase of the desired IQ symbol is created
by adjusting the phase shift of the RIS elements, while a positive
symbol selected from a specific pre-designed PAM constellation forms
the amplitude of that IQ symbol. That is, in the proposed RIS-RQSM
system, in contrast to conventional IQ modulation, the transmitter
constructs the IQ symbol at the receiver with the aid of the RIS elements
and a single radio frequency (RF) chain.
\item We propose an energy-based greedy detector (GD) at the receiver to
detect the indices of the selected antennas with low complexity. Then,
the I and Q symbols can be detected independently by using a one-dimensional
maximum likelihood (ML) detector at each of the detected antennas.
We also propose and design a one-tap zero-forcing (ZF) pre-equalizer
which remarkably reduces the channel state information (CSI) requirement
at the receiver. This yields a significant reduction in the feedback
payload.
\item We analyze the average bit error probability (ABEP) of the proposed
RIS-RQSM system with the GD receiver and derive a closed-form upper
bound which is tight, especially at high SNR values. Then, we propose
an optimization problem to design an IQ modulation scheme in order
to minimize the ABEP. We utilize some accurate approximations to reduce
the complexity of the optimization problem and derive an analytical
solution. Indeed, optimizing the IQ modulation enables the system
to use the GD as an alternative to the ML detector. The results show
that the GD in the RIS-RQSM system with optimized constellation performs
considerably close to the ML detector, such that the performance gap
is negligible.
\item Finally, we compare the bit error rate (BER) performance results with
those of the most prominent benchmark scheme. The results show that
the proposed RIS-RQSM system substantially outperforms the benchmark
scheme. This performance improvement improves with an increasing number
of receive antennas.
\end{itemize}
The rest of this paper is organized as follows. The RIS-RQSM system
model is described in Section~\ref{sec:System-Model}. In Section~\ref{sec:SSK},
we summarize the transceiver design of the RIS-RQSSK system of \cite{dinan2022ris},
which forms the baseline model for the proposed system. The transmitter
and receiver structure design for the proposed RIS-RQSM system is
presented in detail in Section~\ref{sec:SM}. The \acsu{ABEP} performance
of the proposed RIS-RQSM system is analyzed in Section~\ref{sec:Performance-Analysis}.
In Section~\ref{sec:IQ-Modulation-Design}, we formulate the optimization
problem to minimize the system error rate performance and determine
its analytical solution. In Section~\ref{sec:Numerical-Results},
we provide numerical results and comparisons with the benchmark scheme.
Finally, Section~\ref{sec:Conclusion} concludes this paper.

\emph{Notation:} Boldface lower-case letters denote column vectors,
and boldface upper-case letters denote matrices. $\left(\cdot\right)^{\mathcal{R}}$
and $\left(\cdot\right)^{\mathcal{I}}$ denote the real and imaginary
components of a scalar/vector, respectively. $\left(\cdot\right)^{\star}$
represents the optimum value of a scalar/vector variable. $\mathsf{\mathbb{E}}\left\{ \cdot\right\} $
and $\mathbb{V}\left\{ \cdot\right\} $, respectively, denote the
expectation and variance operator. $\mathcal{N}\left(\mu,\sigma^{2}\right)$
(resp., $\mathcal{CN}\left(\mu,\sigma^{2}\right)$) represents the
normal (resp., complex normal) distribution with mean $\mu$ and variance
$\sigma^{2}$. For a real/complex scalar $s$, $\left|s\right|$ denotes
the absolute value, while for a set $\mathcal{S}$, $\left|\mathcal{S}\right|$
denotes its cardinality. $\text{sgn}\left(\cdot\right)$ represents
the sign function which determines the sign of a real variable, i.e.,
for $x\neq0$, it is defined as $\text{sgn}\left(x\right)=\left\{ +1\text{ if }x>0,\,-1\text{ if }x<0\right\} $.
Finally, the set of complex matrices of size $m\times n$ is denoted
by $\mathbb{C}^{m\times n}$.

\section{System Model\label{sec:System-Model}}

\begin{figure}[t]
\begin{centering}
\includegraphics[scale=0.3]{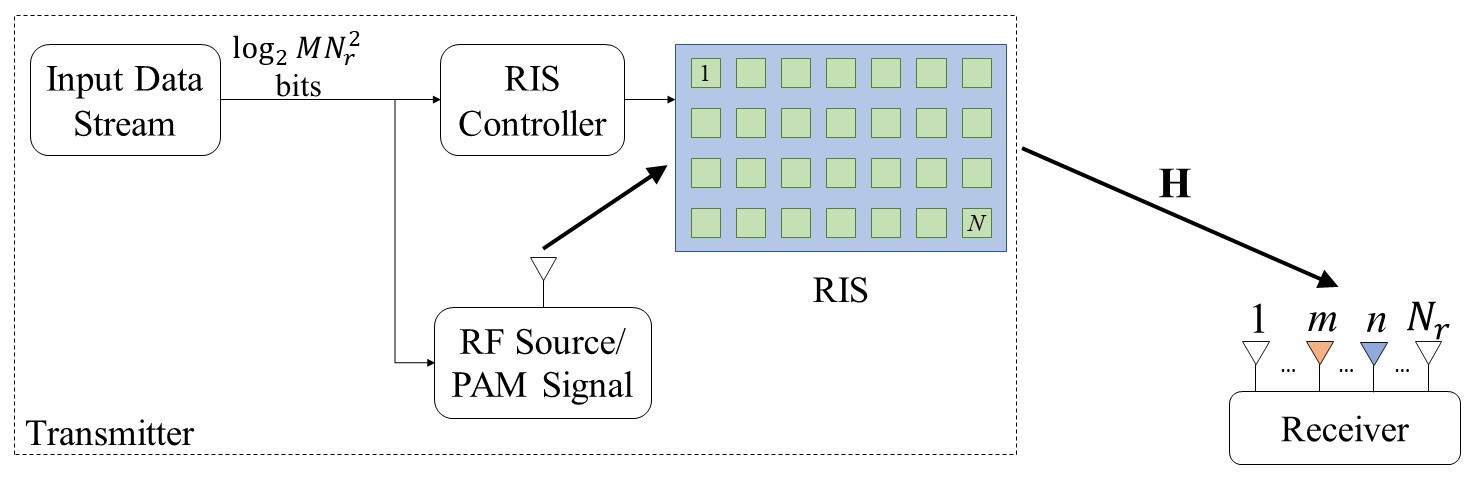}
\par\end{centering}
\caption{A schematic representation of RIS-assisted receive quadrature spatial
modulation (RIS-RQSM) system (in RIS-RQSSK system, an RF source with
constant energy is used).\label{fig:Schematic}}
\end{figure}

In this section, we describe the system model for the proposed RIS-assisted
receive quadrature spatial modulation (RIS-RQSM) scheme. A schematic
of the RIS-RQSM system is presented in Fig.~\ref{fig:Schematic}.
We consider the RIS-AP model \cite{basar2019transmission,basar2019wireless},
where the RIS forms part of the transmitter and reflects the incident
wave emitted from a single transmit antenna which is located in the
vicinity of the \acsu{RIS} such that the path loss and scattering
of the link between the \ac{RIS} and the transmit antenna is negligible.
The RIS is comprised of $N$ reflecting elements whose vector of phase
shifts $\boldsymbol{\theta}\in\mathbb{C}^{N\times1}$ is controlled
by the transmitter to convey information. Here we assume lossless
reflection from the \ac{RIS}, i.e., $\left|\theta_{i}\right|=1$
for $i=1,2,\dots,N$. The receiver is equipped with $N_{r}$ antennas
and is placed far from the transmitter. We assume that the receiver
can only receive the signal reflected from the \ac{RIS} elements
through the wireless fading channel $\mathbf{H}\in\mathbb{C}^{N_{r}\times N}$,
whose elements are assumed to be \ac{iid} according to $\mathcal{CN}\left(0,1\right)$.
In this scenario, the input data stream is split into packets of $\log_{2}MN_{r}^{2}$
bits. The first $\log_{2}N_{r}^{2}$ bits are used to independently
select two receive antennas to convey the spatial symbol, and the
remaining $\log_{2}M$ bits determine the desired IQ symbol that is
selected from an $M$-ary QAM constellation. Unlike in conventional
communication systems, in the RIS-RQSM system the selected IQ symbol
is not transmitted through a single-antenna transmitter, but is created
at the selected receive antennas via both adjusting the RIS phase
shifts \emph{and} emitting a specific PAM symbol from the transmit
antenna\footnote{It is worth mentioning that in contrast to the conventional RIS-SM
system, in the RIS-RQSM the RF source at the transmitter only requires
the hardware for the in-phase (I) signal component, which results
in a lower hardware complexity.}, with a property that the I component appears on the first selected
antenna, while the Q component appears on the second selected antenna.
Thus, the RIS-RQSM scheme represents a significant generalization
of the RIS-assisted receive quadrature space-shift keying (RIS-RQSSK)
system described in \cite{dinan2022ris}. In RIS-RQSSK, an RF source
is used to transmit a constant signal toward the RIS; therefore, only
a spatial symbol can be transmitted, while the PAM signal in RIS-RQSM
enables the transmitter to transfer additional data bits via IQ modulation.
In the next section, we will provide a brief overview of the RIS-RQSSK
system. Then, the proposed RIS-RQSM system will be described in Section~\ref{sec:SM}.

\section{RIS-Assisted Receive Quadrature Space-Shift Keying \cite{dinan2022ris}\label{sec:SSK}}

In this section, we summarize the system model of the RIS-RQSSK scheme
of \cite{dinan2022ris} and outline its phase shift optimization procedure.
In the RIS-RQSSK system, the transmitter is equipped with an RF source
with constant energy $E_{s}$. In this scenario, two receive antennas
are independently selected according to two packets of $\log_{2}N_{r}$
input data bits. Then, the transmitter reflects the signal to the
receiver through the RIS, aiming to simultaneously maximize the \acsu{SNR}
associated to the real part of the signal at the first selected receive
antenna $m$, while also maximizing the \ac{SNR} associated to the
imaginary part of the signal at the second selected receive antenna
$n$. For this system, the real and imaginary components of the baseband
received signal at the selected antennas $m$ and $n$, respectively,
are given by 
\begin{align}
y_{m}^{\mathcal{R}} & =\sqrt{E_{s}}\left[\mathbf{h}_{m}^{\mathcal{R}}\boldsymbol{\theta}^{\mathcal{R}}-\mathbf{h}_{m}^{\mathcal{I}}\boldsymbol{\theta}^{\mathcal{I}}\right]+n_{m}^{\mathcal{R}},\label{eq:real y-SSK}
\end{align}
\begin{align}
y_{n}^{\mathcal{I}} & =\sqrt{E_{s}}\left[\mathbf{h}_{n}^{\mathcal{R}}\boldsymbol{\theta}^{\mathcal{I}}+\mathbf{h}_{n}^{\mathcal{I}}\boldsymbol{\theta}^{\mathcal{R}}\right]+n_{n}^{\mathcal{I}},\label{eq:imag y-SSK}
\end{align}
where $\mathbf{h}_{l}=\left[h_{l,1},h_{l,2},\dots,h_{l,N}\right]$
is the $l$-th row of $\mathbf{H}$, and $n_{l}\in\mathbb{C}$ is
the additive white Gaussian noise at the $l$-th receive antenna that
is distributed according to $\mathcal{CN}\left(0,N_{0}\right)$. To
maximize both SNR components associated to the real and imaginary
parts of the selected receive antennas $m$ and $n$, a \emph{max-min}
optimization problem was defined as 
\begin{align}
\underset{\boldsymbol{\theta}^{\mathcal{R}},\boldsymbol{\theta}^{\mathcal{I}}}{\max} & \ \min\left(\left|\mathbf{h}_{m}^{\mathcal{R}}\boldsymbol{\theta}^{\mathcal{R}}-\mathbf{h}_{m}^{\mathcal{I}}\boldsymbol{\theta}^{\mathcal{I}}\right|,\left|\mathbf{h}_{n}^{\mathcal{R}}\boldsymbol{\theta}^{\mathcal{I}}+\mathbf{h}_{n}^{\mathcal{I}}\boldsymbol{\theta}^{\mathcal{R}}\right|\right)\label{eq: max-min-SSK}\\
\mbox{s.t.}\;\; & \ \left(\theta_{i}^{\mathcal{R}}\right)^{2}+\left(\theta_{i}^{\mathcal{I}}\right)^{2}=1,\ \mbox{for all}\;i=1,2,\dots,N.\nonumber 
\end{align}
Taking the case where the noise-free signal components in (\ref{eq:real y-SSK})
and (\ref{eq:imag y-SSK}) are positive, the optimal values of $\left\{ \theta_{i}^{\mathcal{R}}\right\} $
and $\left\{ \theta_{i}^{\mathcal{I}}\right\} $ are given by 
\begin{equation}
\theta_{i}^{\mathcal{R}\star}=\frac{\lambda A_{i}+\left(1-\lambda\right)B_{i}}{\sqrt{\left(\lambda A_{i}+\left(1-\lambda\right)B_{i}\right)^{2}+\left(\lambda C_{i}+\left(1-\lambda\right)D_{i}\right)^{2}}},\label{eq:optimum theta_R}
\end{equation}
for all $i=1,2,\dots,N$, and 
\begin{equation}
\theta_{i}^{\mathcal{I}\star}=\frac{\lambda C_{i}+\left(1-\lambda\right)D_{i}}{\sqrt{\left(\lambda A_{i}+\left(1-\lambda\right)B_{i}\right)^{2}+\left(\lambda C_{i}+\left(1-\lambda\right)D_{i}\right)^{2}}},\label{eq:optimum theta_I}
\end{equation}
for all $i=1,2,\dots,N$, where we define 
\begin{equation}
A_{i}=h_{m,i}^{\mathcal{R}},\ B_{i}=h_{n,i}^{\mathcal{I}},\ C_{i}=-h_{m,i}^{\mathcal{I}},\ \text{and}\ D_{i}=h_{n,i}^{\mathcal{R}},\label{eq:A,B,C,D_QSSK}
\end{equation}
to simplify the notation, and where, for $N\gg1$, the value of $\lambda\in(0,1)$
is the solution to 
\begin{equation}
f(\lambda)\triangleq\sum_{i=1}^{N}\frac{\left(A_{i}-B_{i}\right)\left(\lambda A_{i}+\left(1-\lambda\right)B_{i}\right)+\left(C_{i}-D_{i}\right)\left(\lambda C_{i}+\left(1-\lambda\right)D_{i}\right)}{\sqrt{\left(\lambda A_{i}+\left(1-\lambda\right)B_{i}\right)^{2}+\left(\lambda C_{i}+\left(1-\lambda\right)D_{i}\right)^{2}}}=0.\label{eq:equation lambda}
\end{equation}
In addition, with the optimal phase shift values given in (\ref{eq:optimum theta_R})
and (\ref{eq:optimum theta_I}), the resulting SNR components have
the same value, i.e., we have 
\[
\mathbf{h}_{m}^{\mathcal{R}}\boldsymbol{\theta}^{\mathcal{R}\star}-\mathbf{h}_{m}^{\mathcal{I}}\boldsymbol{\theta}^{\mathcal{I}\star}=\mathbf{h}_{n}^{\mathcal{R}}\boldsymbol{\theta}^{\mathcal{I}\star}+\mathbf{h}_{n}^{\mathcal{I}}\boldsymbol{\theta}^{\mathcal{R}\star}.
\]

Finally, at the receiver, a simple but effective \acfi{GD} is employed
to detect the selected receive antennas without the need for any knowledge
of the \acsu{CSI} at the receiver. The \ac{GD} operates via 
\begin{align}
\hat{m} & =\arg\max_{m\in\{1,2,\dots,N_{r}\}}\left\{ \left(y_{m}^{\mathcal{R}}\right)^{2}\right\} ,\label{eq:GD m_hat}\\
\hat{n} & =\arg\max_{n\in\{1,2,\dots,N_{r}\}}\left\{ \left(y_{n}^{\mathcal{I}}\right)^{2}\right\} .\label{eq:GD n_hat}
\end{align}

The performance results have demonstrated the superiority of the RIS-RQSSK
system over comparable benchmark schemes. This motivates us to extend
this scheme to the context of QSM, which is the subject of the next
section.

\section{RIS-Assisted Receive Quadrature Spatial Modulation\label{sec:SM}}

In general, while the spectral efficiency of an SSK system can be
increased by extending it to the corresponding quadrature SSK system,
it can be further improved by implementing a conventional IQ modulation
on top of the antenna index modulation. In the conventional receive
quadrature SM (RQSM), the transmit vector can be designed to place
the real and imaginary parts of the symbol separately at a specific
position of the real and imaginary receive vector. On the other hand,
in the RIS-RQSM scheme, the transmitter is equipped with only one
antenna and therefore can only transmit one symbol in each symbol
interval. In addition, since the real and imaginary parts of the desired
symbol needs to be separated at the receiver, the transmitter can
only perform amplitude modulation through the RF source to be detectable
at the receiver (as also suggested in \cite{yuan2021receive} for
the RIS-RQRM scheme), i.e., it is not feasible to transmit a QAM symbol
and receive the I and Q components separately at two different receive
antennas. To tackle this problem, in the proposed RIS-RQSM system
we introduce a new paradigm in order to \emph{construct} an $M$-ary
QAM symbol (in fact, two independent symbols from identical $\sqrt{M}$-ary
PAM constellations) at the receiver via the adjustment of both the
amplitude of the RF source and the phase shifts of the RIS elements.
Therefore, in the RIS-RQSM system the rate is $R=\log_{2}M+2\log_{2}N_{r}$
\ac{bpcu}. In this scenario, the desired received signal components
are given by 
\begin{align}
y_{m}^{\mathcal{R}} & =\left[\mathbf{h}_{m}^{\mathcal{R}}\boldsymbol{\theta}^{\mathcal{R}}-\mathbf{h}_{m}^{\mathcal{I}}\boldsymbol{\theta}^{\mathcal{I}}\right]Gs+n_{m}^{\mathcal{R}},\label{eq:real y-SM}
\end{align}
\begin{align}
y_{n}^{\mathcal{I}} & =\left[\mathbf{h}_{n}^{\mathcal{R}}\boldsymbol{\theta}^{\mathcal{I}}+\mathbf{h}_{n}^{\mathcal{I}}\boldsymbol{\theta}^{\mathcal{R}}\right]Gs+n_{n}^{\mathcal{I}},\label{eq:imag y-SM}
\end{align}
where $s$ is the transmit symbol selected from a specific positive
real PAM constellation, denoted by $\mathcal{P}_{\mathrm{RF}}$. The
amplitudes in $\mathcal{P}_{\mathrm{RF}}$ are the magnitudes of the
complex symbols in an $M$-ary QAM constellation $\mathcal{M}$ with
average energy $E_{s}$, i.e., $s=\left|x\right|$ where $x\in\mathcal{M}$
is the desired IQ symbol, and $G>0$ is a one-tap zero-forcing (ZF)
pre-equalizer to be defined later.

To produce the desired $M$-ary QAM signal at the receiver, we modify
the problem in (\ref{eq: max-min-SSK}) to accommodate both the index
modulation and IQ modulation as \begin{subequations}
\label{eq: max-min-SM}
\begin{align}
\underset{\boldsymbol{\theta}^{\mathcal{R}},\boldsymbol{\theta}^{\mathcal{I}}}{\max}&\ \min\left(Y_{R},\delta Y_{I}\right)\\
\mbox{s.t.}\;\;&\;Y_{R}=\text{sgn}\left(x^{\mathcal{R}}\right)\left(\mathbf{h}_{m}^{\mathcal{R}}\boldsymbol{\theta}^{\mathcal{R}}-\mathbf{h}_{m}^{\mathcal{I}}\boldsymbol{\theta}^{\mathcal{I}}\right),\\
&\;Y_{I}=\text{sgn}\left(x^{\mathcal{I}}\right)\left(\mathbf{h}_{n}^{\mathcal{R}}\boldsymbol{\theta}^{\mathcal{I}}+\mathbf{h}_{n}^{\mathcal{I}}\boldsymbol{\theta}^{\mathcal{R}}\right),\\
&\ \left(\theta_{i}^{\mathcal{R}}\right)^{2}+\left(\theta_{i}^{\mathcal{I}}\right)^{2}=1,\ \mbox{for all}\;i=1,2,\dots,N,
\end{align}
\end{subequations}where $\delta>0$ is the absolute value of the ratio of the real to
the imaginary part of $x$, i.e., $\delta=\left|x^{\mathcal{R}}/x^{\mathcal{I}}\right|$.
It can be seen that this optimization problem is similar to the optimization
problem for the RIS-RQSSK scenario; hence, it can be solved by a similar
approach to that used in \cite{dinan2022ris} (we omit the details
for brevity). As a result, $\left\{ \theta_{i}^{\mathcal{R}\star}\right\} $
and $\left\{ \theta_{i}^{\mathcal{I}\star}\right\} $ are again given
by (\ref{eq:optimum theta_R}) and (\ref{eq:optimum theta_I}), respectively,
and $\lambda$ can also be evaluated by solving (\ref{eq:equation lambda});
however, it is required to re-define the variables in (\ref{eq:A,B,C,D_QSSK})
accordingly as 
\begin{equation}
A_{i}=\text{sgn}\left(x^{\mathcal{R}}\right)h_{m,i}^{\mathcal{R}},\ B_{i}=\delta\,\text{sgn}\left(x^{\mathcal{I}}\right)h_{n,i}^{\mathcal{I}},\ C_{i}=\text{sgn}\left(x^{\mathcal{R}}\right)\left(-h_{m,i}^{\mathcal{I}}\right),\ \text{and}\ D_{i}=\delta\,\text{sgn}\left(x^{\mathcal{I}}\right)h_{n,i}^{\mathcal{R}}.\label{eq:A,B,C,D_QSM}
\end{equation}
Note that the maximization problem forces $Y_{R}$ and $Y_{I}$ to
be positive. As a result, the sign functions in (\ref{eq: max-min-SM})
determine the signs of the noise-free received signal components.
To elucidate the functionality of the optimization problem above,
we take symbol $x=1-3j$ as an example; then, we have $\text{sgn}\left(x^{\mathcal{R}}=1\right)=+1$
and $\text{sgn}\left(x^{\mathcal{I}}=-3\right)=-1$. Therefore, we
obtain $Y_{R}=+\left(\mathbf{h}_{m}^{\mathcal{R}}\boldsymbol{\theta}^{\mathcal{R}}-\mathbf{h}_{m}^{\mathcal{I}}\boldsymbol{\theta}^{\mathcal{I}}\right)>0$
and $Y_{I}=-\left(\mathbf{h}_{n}^{\mathcal{R}}\boldsymbol{\theta}^{\mathcal{I}}+\mathbf{h}_{n}^{\mathcal{I}}\boldsymbol{\theta}^{\mathcal{R}}\right)>0$,
which indicates that the real component of the constructed received
symbol is positive and its imaginary component is negative, similar
to the selected symbol $x$. It is also worth pointing out that at
the optimal point, the values involved in the minimization are equal,
i.e., with the values $\left\{ \theta_{i}^{\mathcal{R}\star}\right\} $
and $\left\{ \theta_{i}^{\mathcal{I}\star}\right\} $ we have $Y_{R}^{\star}=\delta Y_{I}^{\star}$,
where $Y_{R}^{\star}$ and $Y_{I}^{\star}$ are the optimum values
of $Y_{R}$ and $Y_{I}$ produced by (\ref{eq: max-min-SM}). Hence,
we can conclude that the \emph{phase} of the desired QAM symbol is
correctly designed. Next, in order to explain why the PAM constellation
$\mathcal{P}_{\mathrm{RF}}$ must be utilized at the transmitter,
we need to ascertain how the RIS-aided channel acts for various values
of $\delta$.

Due to the presence of random variables in (\ref{eq:equation lambda}),
$\lambda$ also presents a random behavior. It is not easy to determine
the stochastic characteristics (e.g., mean and variance) of $\lambda$
from (\ref{eq:equation lambda}); however, experimental results provide
strong evidence that the mean value of $\lambda$ is $\mathbb{E}\left\{ \lambda\right\} =\bar{\lambda}=\frac{\delta^{2}}{1+\delta^{2}}$
and that its variance tends to zero with an increasing number of RIS
elements $N$. This observation can be further used to approximate
the average value of the optimum objective in (\ref{eq: max-min-SM}),
which is provided in the following theorem.
\begin{thm}
For large values of $N$, the means $\mathbb{E}\left\{ Y_{R}^{\star}\right\} $
and $\mathbb{E}\left\{ Y_{I}^{\star}\right\} $ can be closely approximated
by \label{thm:ave_Y}
\[
\mathbb{E}\left\{ Y_{R}^{\star}\right\} \approx\sqrt{\bar{\lambda}}\frac{N\sqrt{\pi}}{2},\ \mathbb{E}\left\{ Y_{I}^{\star}\right\} \approx\sqrt{1-\bar{\lambda}}\frac{N\sqrt{\pi}}{2}.
\]
\end{thm}
\begin{IEEEproof}
The proof is provided in Appendix \ref{sec:proof ave}.
\end{IEEEproof}
From Theorem~\ref{thm:ave_Y}, it can be observed that the mean value
of the complex symbol created by the received signal components at
the selected antennas lies on a circle with radius $\beta=\frac{N\sqrt{\pi}}{2}$
for any value of $\delta$. Therefore, in addition to optimizing the
phase angles of the RIS elements, an appropriate positive PAM symbol
$s\in\mathcal{P}_{\mathrm{RF}}$ is required to be modulated at the
RF source in order to adjust the magnitude of the received signal
to accommodate the desired QAM symbol in a predefined constellation.
In other words, the phase of the QAM symbol is determined by the RIS
elements while its amplitude is determined by the PAM symbol. Therefore,
the transmit symbol $s=\left|x\right|$ is required at the RF source.

The symbol $s$ is then multiplied by $G$ at the transmitter to ensure
that the gain of the link is constant at all times (i.e., for each
symbol and for each channel realization). Therefore, we design $G$
via 
\begin{equation}
G=\frac{\mathbb{E}\left\{ Y_{R}^{\star}\right\} }{Y_{R}^{\star}}=\frac{\mathbb{E}\left\{ \mathbf{h}_{m}^{\mathcal{R}}\boldsymbol{\theta}^{\mathcal{R}\star}-\mathbf{h}_{m}^{\mathcal{I}}\boldsymbol{\theta}^{\mathcal{I}\star}\right\} }{\mathbf{h}_{m}^{\mathcal{R}}\boldsymbol{\theta}^{\mathcal{R}\star}-\mathbf{h}_{m}^{\mathcal{I}}\boldsymbol{\theta}^{\mathcal{I}\star}},\label{eq:G}
\end{equation}
where $\boldsymbol{\theta}^{\star}$ is the optimum vector of phase
shifts of the RIS elements associated to the desired transmit symbol.
Note that $G$ has a value that is specific to each symbol $x$ and
channel realization $\mathbf{H}$. In fact, $G$ can be realized as
a one-tap ZF pre-equalizer. As a result, the receiver only needs to
know the \emph{effective gain} of the RIS-assisted wireless channel,
i.e, the gain of the equivalent Gaussian channel which is obtained
by the aid of the RIS elements, which is equal to $\beta^{2}$; no
additional CSI is necessary for the GD detector, which significantly
reduces the feedback payload of the system. On the other hand, the
CSI must be available at the transmitter in order to adjust the phase
shifts of the RIS elements and implement the one-tap pre-equalizer.
\begin{thm}
Under the assumption of a large number of RIS elements, the mean values
of $G$ and $G^{2}$ both tend to unity, i.e., $\underset{N\rightarrow\infty}{\lim}\mathbb{E}\left\{ G\right\} =1$
and $\underset{N\rightarrow\infty}{\lim}\mathbb{E}\left\{ G^{2}\right\} =1$.\label{thm:ave_G2}
\end{thm}
\begin{IEEEproof}
Here we only prove that $\underset{N\rightarrow\infty}{\lim}\mathbb{E}\left\{ G^{2}\right\} =1$.
The convergence of the mean value of $G$ can be derived in a similar
manner. The mean value of $G^{2}$ is given by 
\[
\mathbb{E}\left\{ G^{2}\right\} =\mathbb{E}\left\{ \frac{\mathbb{E}^{2}\left\{ Y_{R}^{\star}\right\} }{\left(Y_{R}^{\star}\right)^{2}}\right\} =\mu^{2}\mathbb{E}\left\{ \frac{1}{\left(Y_{R}^{\star}\right)^{2}}\right\} ,
\]
where $\mu=\mathbb{E}\left\{ Y_{R}^{\star}\right\} =\sqrt{\bar{\lambda}}\frac{N\sqrt{\pi}}{2}$.
According to the \ac{CLT}, $Y_{R}^{\star}$ is distributed as\footnote{This is proved in \cite{dinan2022ris} for the RIS-RQSSK scenario,
i.e., for $\delta=1$, however, the proof can be extended to the general
case where $\delta=\left|x^{\mathcal{R}}/x^{\mathcal{I}}\right|$
(for brevity, these details are omitted). Later (in Section~\ref{sec:Performance-Analysis})
we will show how the variance $\sigma^{2}$ is related to $N$.} $Y_{R}^{\star}\sim\mathcal{N}\left(\mu,\sigma^{2}\right)$, where
$\sigma^{2}\propto N$. Then, the average of $G^{2}$ can be expressed
as 
\begin{align*}
\mathbb{E}\left\{ G^{2}\right\}  & =\mu^{2}\mathbb{E}\left\{ \frac{1}{\left(Y_{R}^{\star}\right)^{2}}\right\} =\frac{\mu^{2}}{\sqrt{2\pi\sigma^{2}}}\int_{-\infty}^{\infty}\frac{1}{y^{2}}e^{-\frac{\left(y-\mu\right)^{2}}{2\sigma^{2}}}\mathrm{d}y\\
 & =\frac{1}{\sqrt{2\pi}}\frac{\mu^{2}}{\sigma^{2}}\int_{-\infty}^{\infty}\frac{1}{\left(u+\frac{\mu}{\sigma}\right)^{2}}e^{-\frac{u^{2}}{2}}\mathrm{d}u,
\end{align*}
where we used the change of variable $u=\frac{y-\mu}{\sigma}$. Since
$\frac{\mu}{\sigma}\propto\sqrt{N}\rightarrow\infty$ as $N\rightarrow\infty$,
we can write
\[
\lim_{N\to\infty}\mathbb{E}\left\{ G^{2}\right\} =\lim_{\frac{\mu}{\sigma}\to\infty}\frac{1}{\sqrt{2\pi}}\frac{\mu^{2}}{\sigma^{2}}\int_{-\infty}^{\infty}\frac{1}{\left(u+\frac{\mu}{\sigma}\right)^{2}}e^{-\frac{u^{2}}{2}}\mathrm{d}u=\frac{1}{\sqrt{2\pi}}\int_{-\infty}^{\infty}e^{-\frac{u^{2}}{2}}\mathrm{d}u=1.
\]
\end{IEEEproof}
Note that in practice, the number of RIS elements is large enough
so that the expressions in Theorem~\ref{thm:ave_G2} serve as accurate
approximations for our design. Theorem~\ref{thm:ave_G2} implies that
the pre-equalizer $G$ does not change the average transmit power
of the system, i.e., $\mathbb{E}\left\{ \left(Gs\right)^{2}\right\} =E_{s}$;
hence the SNR is simply given by $E_{s}/N_{0}$.

\subsection*{Receiver Structure}

Similar to the RIS-RQSSK scheme, the receiver can employ a GD to detect
the selected antenna indices via (\ref{eq:GD m_hat}) and (\ref{eq:GD n_hat}).
After this, the receiver can demodulate the desired I and Q symbols
via 

\begin{equation}
\hat{x}^{\mathcal{R}}=\arg\underset{x^{\mathcal{R}}}{\min}\left\{ \left(y_{m}^{\mathcal{R}}-\beta x^{\mathcal{R}}\right)^{2}\right\} ,\label{eq:x_hat-real}
\end{equation}
\begin{equation}
\hat{x}^{\mathcal{I}}=\arg\underset{x^{\mathcal{I}}}{\min}\left\{ \left(y_{n}^{\mathcal{I}}-\beta x^{\mathcal{I}}\right)^{2}\right\} ,\label{eq:x_hat-imag}
\end{equation}
where $\beta=\frac{N\sqrt{\pi}}{2}$ is the \emph{effective} channel
coefficient.

On the other hand, the \ac{ML} detector for the proposed RIS-RQSM
system operates via 

\begin{equation}
\left(\hat{m},\hat{n},\hat{x}\right)=\arg\min_{m,n,x}\sum_{l=1}^{N_{r}}\left(y_{l}-\mathbf{h}_{l}\boldsymbol{\theta}^{\star}Gs\right)^{2},\label{eq:ML}
\end{equation}
where we note that $\boldsymbol{\theta}^{\star}$ is a multi-variable
function of $(m,n,x)$, and $s=\left|x\right|$. While the GD is CSI-free,
the ML detector relies on having full CSI at the receiver. Furthermore,
it can be seen that the ML detector needs to compute $\boldsymbol{\theta}^{\star}$
for all combinations of the selected receive antennas and then search
over all possible combinations of the spatial symbols and IQ modulation
symbols. These facts make the ML detector significantly more complex
than GD. Although the ML detector provides an optimum receiver, we
will show later in Sections~\ref{sec:IQ-Modulation-Design}~and~\ref{sec:Numerical-Results}
that optimizing the IQ constellation, in addition to increasing the
performance of the system, can also leverage the GD efficiency such
that it competes very strongly with the ML detector (i.e., the performance
gap is negligible).

\section{Performance Analysis\label{sec:Performance-Analysis}}

In this section, we analyze the \ac{ABEP} of the proposed RIS-RQSM
system. This analysis focuses on the GD receiver. Here we only perform
the analysis for the detection of the antenna $m$ with active real
part along with the real part of the corresponding modulated IQ symbol,
$x^{\mathcal{R}}$; due to the inherent symmetry in the expressions,
it is easy to show that the ABEP expression for the detection of the
antenna $n$ with active imaginary part along with the imaginary part
of the corresponding modulated IQ symbol $x^{\mathcal{I}}$ is identical.
An upper bound on the ABEP, which is tight especially at high SNR,
is given by 
\begin{equation}
\mathrm{ABEP}\leq\frac{1-P_{e}\left(m\right)}{\sqrt{M}\log_{2}\left(\sqrt{M}N_{r}\right)}\sum_{x^{\mathcal{R}}}\sum_{\hat{x}^{\mathcal{R}}\neq x^{\mathcal{R}}}\mathrm{PEP}\left(x^{\mathcal{R}}\rightarrow\hat{x}^{\mathcal{R}}|m=\hat{m}\right)e\left(x^{\mathcal{R}}\rightarrow\hat{x}^{\mathcal{R}}\right)+0.5P_{e}\left(m\right),\label{eq:ABEP}
\end{equation}
where $P_{e}\left(m\right)$ is the probability of erroneous detection
of the selected receive antenna $m$, $\mathrm{PEP}\left(x^{\mathcal{R}}\rightarrow\hat{x}^{\mathcal{R}}|m=\hat{m}\right)$
is the \ac{PEP} associated with the real part of the symbols $x$
and $\hat{x}$ conditioned on correct detection of the antenna index,
and $e\left(x^{\mathcal{R}}\rightarrow\hat{x}^{\mathcal{R}}\right)$
is the Hamming distance between the binary representations of the
real parts of the symbols $x$ and $\hat{x}$. Here we assume that
half of the bits are in error under the condition of erroneous index
detection (note that this assumption represents the worst-case scenario),
so that $P_{e}\left(m\right)$ can be written as 
\begin{equation}
P_{e}\left(m\right)=\left(N_{r}-1\right)\overline{\mathrm{PEP}}\left(m\rightarrow\hat{m}\right),\label{eq:P_e(m)}
\end{equation}
where $\overline{\mathrm{PEP}}\left(m\rightarrow\hat{m}\right)$ is
the average PEP associated with the antenna indices $m$ and $\hat{m}$,
and is given by 
\begin{align}
\overline{\mathrm{PEP}}\left(m\rightarrow\hat{m}\right) & =\frac{1}{\sqrt{M}}\sum_{x^{\mathcal{R}}}\mathrm{PEP}\left(m\rightarrow\hat{m}|x^{\mathcal{R}}\right)\nonumber \\
 & =\frac{1}{\sqrt{M}}\sum_{x^{\mathcal{R}}\in\mathcal{M}_{R}}\frac{2}{\sqrt{M}}\sum_{\delta\in\mathcal{D}_{x^{\mathcal{R}}}}\mathrm{PEP}\left(m\rightarrow\hat{m}|x^{\mathcal{R}},\delta\right),\label{eq:ave_PEP}
\end{align}
where $\mathcal{M}_{R}$ is the set consisting of all possible values
of $x^{\mathcal{R}}$, the real component of symbols in $\mathcal{M}$,
with $\left|\mathcal{M}_{R}\right|=\sqrt{M}$, and $\mathcal{D_{\xi}}=\left\{ \left|\frac{x^{\mathcal{R}}}{x^{\mathcal{I}}}\right||x^{\mathcal{R}}=\xi,x^{\mathcal{I}}\in\mathcal{M}_{I}\right\} $
with $\left|\mathcal{D}_{\xi}\right|=\frac{\sqrt{M}}{2}$ (where $\mathcal{M}_{I}$
is the set consisting of all possible values of $x^{\mathcal{I}}$);
for instance, for a conventional 16-QAM constellation we have $\mathcal{M}_{R}=\mathcal{M}_{I}=\left\{ -3,-1,1,3\right\} $,
and for $x^{\mathcal{R}}=\left\{ -1,1\right\} $ we have $\mathcal{D}_{-1}=\mathcal{D}_{1}=\left\{ 1,1/3\right\} $,
while for $x^{\mathcal{R}}=\left\{ -3,3\right\} $ we have $\mathcal{D}_{-3}=\mathcal{D}_{3}=\left\{ 1,3\right\} $.
Considering the use of GD at the receiver, the PEP associated with
the selected antenna $m$ and the detected antenna $\hat{m}\ne m$
conditioned on the selected symbol $x$ (i.e., given $x^{\mathcal{R}}$
and $\delta$) is given by 
\begin{align}
\mathrm{PEP}\left(m\rightarrow\hat{m}|x^{\mathcal{R}},\delta\right)= & \Pr\left\{ \left(y_{m}^{\mathcal{R}}\right)^{2}<\left(y_{\hat{m}}^{\mathcal{R}}\right)^{2}|x^{\mathcal{R}},\delta\right\} \nonumber \\
= & \Pr\left\{ \left(\left[\mathbf{h}_{m}^{\mathcal{R}}\boldsymbol{\theta}^{\mathcal{R}\star}-\mathbf{h}_{m}^{\mathcal{I}}\boldsymbol{\theta}^{\mathcal{I}\star}\right]Gs+n_{m}^{\mathcal{R}}\right)^{2}\right.\nonumber \\
 & \left.<\left(\left[\mathbf{h}_{\hat{m}}^{\mathcal{R}}\boldsymbol{\theta}^{\mathcal{R}\star}-\mathbf{h}_{\hat{m}}^{\mathcal{I}}\boldsymbol{\theta}^{\mathcal{I}\star}\right]Gs+n_{\hat{m}}^{\mathcal{R}}\right)^{2}|x^{\mathcal{R}},\delta\right\} \nonumber \\
\approx & \Pr\left\{ \left|Z_{1}\right|<\left|Z_{2}\right|\right\} ,\label{eq:PEP}
\end{align}
where we define $Z_{1}\triangleq\left[\mathbf{h}_{m}^{\mathcal{R}}\boldsymbol{\theta}^{\mathcal{R}\star}-\mathbf{h}_{m}^{\mathcal{I}}\boldsymbol{\theta}^{\mathcal{I}\star}\right]\frac{\left|x^{\mathcal{R}}\right|}{\sqrt{\bar{\lambda}}}+n_{m}^{\mathcal{R}}$
and $Z_{2}\triangleq\left[\mathbf{h}_{\hat{m}}^{\mathcal{R}}\boldsymbol{\theta}^{\mathcal{R}\star}-\mathbf{h}_{\hat{m}}^{\mathcal{I}}\boldsymbol{\theta}^{\mathcal{I}\star}\right]\frac{\left|x^{\mathcal{R}}\right|}{\sqrt{\bar{\lambda}}}+n_{\hat{m}}^{\mathcal{R}}$,
and we have used the approximations stated in Theorem~\ref{thm:ave_G2},
i.e., $\mathbb{E}\left\{ G\right\} \approx1$ and $\mathbb{V}\left\{ G\right\} =\mathbb{E}\left\{ G^{2}\right\} -\mathbb{E}\left\{ G\right\} ^{2}\approx0$,
and we know that $s=\frac{\left|x^{\mathcal{R}}\right|}{\sqrt{\bar{\lambda}}}$,
since $\bar{\lambda}=\frac{\delta^{2}}{1+\delta^{2}}$. To calculate
the probability above, the distributions of $Z_{1}$, in the cases
where $m=n$ and $m\neq n$, and $Z_{2}$, in the cases where $\hat{m}=n$
and $\hat{m}\neq n$, are required. In \cite[Theorems 1-3]{dinan2022ris},
the distributions of the random variables (RVs) $Z_{1}$ and $Z_{2}$
were derived for the case of RIS-RQSSK (in that case it was shown
that $\bar{\lambda}=1/2$). The distributions of $Z_{1}$ and $Z_{2}$
for the more general case of RIS-RQSM can be derived in a similar
manner (we omit the details for brevity).

In the case where $m=n$, with reference to the CLT, $Z_{1}$ is approximately
distributed according to $\mathcal{N}\left(\mu_{1},\sigma_{1}^{2}\right)$,
where $\mu_{1}=\frac{N\sqrt{\pi}}{2}x^{\mathcal{R}}$ and $\sigma_{1}^{2}=N\left(x^{\mathcal{R}}\right)^{2}\frac{4-\pi}{4}+\frac{N_{0}}{2}$.
In the case where $m\neq n$, the mean $\mu_{1}$ is given by the
same expression as in the case where $m=n$, and experimental results
provide strong evidence that the variance of $Z_{1}$ is also exactly
the same as in the case where $m=n$.

On the other hand, $Z_{2}$ is approximately distributed according
to $\mathcal{N}\left(0,\sigma_{2}^{2}\right)$, where the variance
in each case of $\hat{m}=n$ and $\hat{m}\neq n$ is given by 

1) $\hat{m}\neq n$: 
\begin{equation}
\sigma_{2}^{2}=\rho_{1}^{2}\triangleq\frac{N\left(x^{\mathcal{R}}\right)^{2}}{2\bar{\lambda}}+\frac{N_{0}}{2},\label{eq:rho_1^2}
\end{equation}

2) $\hat{m}=n$:
\begin{equation}
\sigma_{2}^{2}=\rho_{2}^{2}\triangleq\frac{N\left(x^{\mathcal{R}}\right)^{2}}{2}+\frac{N_{0}}{2}.\label{eq:rho_2^2}
\end{equation}
Therefore, to calculate the PEP, two different events need to be taken
into consideration: i) $\left\{ \mathcal{E}_{1}:m,\hat{m}\in\left\{ 1,2,\dots,N_{r}\right\} ,\ \hat{m}\neq n\right\} $,
and ii) $\left\{ \mathcal{E}_{2}:m\in\left\{ 1,2,\dots,N_{r}\right\} ,\ \hat{m}=n\right\} $.

It is worth pointing out that $Z_{1}$ and $Z_{2}$ represent the
real part of the signal received at the selected antenna $m$ (having
mean $\mu_{1}\propto N\gg1$) and at a non-selected antenna $\hat{m}$
(having mean zero), respectively. This is the reason that the GD is
able to easily detect the index of the selected receive antenna.

Next, we consider the instance where $x^{\mathcal{R}}>0$ (it is clear
that the PEP for $x^{\mathcal{R}}<0$ is the same). Considering the
distribution of $Z_{1}$, it can be seen that $\frac{\mu_{1}}{\sigma_{1}}\propto\sqrt{N}$
for relatively high SNR values, so that $\frac{\mu_{1}}{\sigma_{1}}\gg1$;
as a result, we have $Z_{1}>0$ with extremely high probability. Hence,
the PEP can be written as 
\begin{align*}
\mathrm{PEP}\left(m\rightarrow\hat{m}|x^{\mathcal{R}},\delta\right)= & \mathrm{PEP}\left(m\rightarrow\hat{m}|x^{\mathcal{R}}>0,\delta\right)\\
\approx & \frac{N_{r}-1}{N_{r}}\Pr\left\{ Z_{1}<\left|Z_{2}\right||\mathcal{E}_{1}\right\} +\frac{1}{N_{r}}\Pr\left\{ Z_{1}<\left|Z_{2}\right||\mathcal{E}_{2}\right\} \\
= & \frac{N_{r}-1}{N_{r}}\int_{0}^{\infty}\Pr\left\{ Z_{1}=\alpha,\left|Z_{2}\right|>\alpha|\mathcal{E}_{1}\right\} \mathrm{d}\alpha\\
 & +\frac{1}{N_{r}}\int_{0}^{\infty}\Pr\left\{ Z_{1}=\alpha,\left|Z_{2}\right|>\alpha|\mathcal{E}_{2}\right\} \mathrm{d}\alpha.
\end{align*}
The above two integrals can be evaluated in a unified manner via 
\begin{align}
\mathrm{I}_{i}\triangleq & \int_{0}^{\infty}\Pr\left\{ Z_{1}=\alpha,\left|Z_{2}\right|>\alpha|\mathcal{E}_{i}\right\} \mathrm{d}\alpha=2\int_{0}^{\infty}p_{z_{1}|\mathcal{E}_{i}}\left(\alpha\right)\Pr\left\{ Z_{2}>\alpha|\mathcal{E}_{i}\right\} \mathrm{d}\alpha\nonumber \\
= & \frac{\sqrt{2}}{\sigma_{1}\sqrt{\pi}}\int_{0}^{\infty}e^{-\frac{1}{2}\left(\frac{\mu_{1}-\alpha}{\sigma_{1}}\right)^{2}}\mathrm{Q}\left(\frac{\alpha}{\rho_{i}}\right)\mathrm{d}\alpha,\;i=1,2.\label{eq:I1}
\end{align}
Applying the exponential approximation of the Q-function as $\mathrm{Q}\left(x\right)\approx\frac{1}{12}e^{-\frac{x^{2}}{2}}+\frac{1}{4}e^{-\frac{2x^{2}}{3}}$
from \cite{chiani2003capacity}, $\mathrm{I}_{i}$ is approximately
given by 
\[
\mathrm{I}_{i}\approx\frac{\sqrt{2}}{\sigma_{1}\sqrt{\pi}}\int_{0}^{\infty}e^{-\frac{1}{2}\left(\frac{\mu_{1}-\alpha}{\sigma_{1}}\right)^{2}}\left[\frac{1}{12}e^{-\frac{1}{2}\left(\frac{\alpha}{\rho_{i}}\right)^{2}}+\frac{1}{4}e^{-\frac{2}{3}\left(\frac{\alpha}{\rho_{i}}\right)^{2}}\right]\mathrm{d}\alpha,\;i=1,2.
\]
After some manipulations we obtain 
\begin{align}
\mathrm{I}_{i} & \approx\frac{\sqrt{2}}{\sigma_{1}\sqrt{\pi}}\left[\frac{1}{12}e^{u_{0,i}}\int_{0}^{\infty}e^{-\frac{1}{2}\left(\frac{\alpha-m_{0,i}}{s_{0,i}}\right)^{2}}\mathrm{d}\alpha+\frac{1}{4}e^{u_{1,i}}\int_{0}^{\infty}e^{-\frac{1}{2}\left(\frac{\alpha-m_{1,i}}{s_{1,i}}\right)^{2}}\mathrm{d}\alpha\right]\nonumber \\
 & =\frac{1}{\sigma_{1}}\left[\frac{1}{6}e^{u_{0,i}}s_{0,i}\mathrm{Q}\left(-\frac{m_{0,i}}{s_{0,i}}\right)+\frac{1}{2}e^{u_{1,i}}s_{1,i}\mathrm{Q}\left(-\frac{m_{1,i}}{s_{1,i}}\right)\right],\;i=1,2,\label{eq:I1-approx}
\end{align}
where 
\[
u_{0,i}=-\frac{1}{2}\frac{\mu_{1}^{2}}{\sigma_{1}^{2}+\rho_{i}^{2}},\;s_{0,i}=\frac{\sigma_{1}\rho_{i}}{\sqrt{\sigma_{1}^{2}+\rho_{i}^{2}}},\;m_{0,i}=\frac{\mu_{1}\rho_{i}^{2}}{\sigma_{1}^{2}+\rho_{i}^{2}},
\]
\[
u_{1,i}=-\frac{2}{3}\frac{\mu_{1}^{2}}{\frac{4}{3}\sigma_{1}^{2}+\rho_{i}^{2}},\;s_{1,i}=\frac{\sigma_{1}\rho_{i}}{\sqrt{\frac{4}{3}\sigma_{1}^{2}+\rho_{i}^{2}}},\;m_{1,i}=\frac{\mu_{1}\rho_{i}^{2}}{\frac{4}{3}\sigma_{1}^{2}+\rho_{i}^{2}},\;i=1,2.
\]
It is easy to see that $\frac{m_{0,i}}{s_{0,i}}$ and $\frac{m_{1,i}}{s_{1,i}}$,
$i=1,2$, have relatively large values for large $N$, such that the
approximations $\mathrm{Q}\left(-\frac{m_{0,i}}{s_{0,i}}\right)\approx\mathrm{Q}\left(-\frac{m_{1,i}}{s_{1,i}}\right)\approx1$
are very accurate; therefore, $\mathrm{I}_{i}$ can be written as
\begin{equation}
\mathrm{I}_{i}\approx\frac{\rho_{i}}{6\sqrt{\sigma_{1}^{2}+\rho_{i}^{2}}}e^{-\frac{1}{2}\frac{\mu_{1}^{2}}{\sigma_{1}^{2}+\rho_{i}^{2}}}+\frac{\rho_{i}}{2\sqrt{\frac{4}{3}\sigma_{1}^{2}+\rho_{i}^{2}}}e^{-\frac{2}{3}\frac{\mu_{1}^{2}}{\frac{4}{3}\sigma_{1}^{2}+\rho_{i}^{2}}},\;i=1,2.\label{eq:I1-approx-Q=00003D1}
\end{equation}
Hence, $P_{e}\left(m\right)$ is given by 
\begin{equation}
P_{e}\left(m\right)=\frac{2\left(N_{r}-1\right)}{M}\sum_{x^{\mathcal{R}}}\sum_{\delta}\left(\frac{N_{r}-1}{N_{r}}\mathrm{I}_{1}+\frac{1}{N_{r}}\mathrm{I}_{2}\right).\label{eq:Pe(m)-final}
\end{equation}
Finally, $\mathrm{PEP}\left(x^{\mathcal{R}}\rightarrow\hat{x}^{\mathcal{R}}|m=\hat{m}\right)$
can be expressed as 
\begin{equation}
\mathrm{PEP}\left(x^{\mathcal{R}}\rightarrow\hat{x}^{\mathcal{R}}|m=\hat{m}\right)=\mathrm{Q}\left(\sqrt{\frac{\beta^{2}\left(x^{\mathcal{R}}-\hat{x}^{\mathcal{R}}\right)^{2}}{2N_{0}}}\right).\label{eq:PEP_IQ}
\end{equation}
Substituting (\ref{eq:Pe(m)-final}) and (\ref{eq:PEP_IQ}) into (\ref{eq:ABEP}),
an accurate closed-form approximation for the ABEP of the RIS-RQSM
system can be obtained.

\section{IQ Modulation Design\label{sec:IQ-Modulation-Design}}

A significant advantage of the proposed RQSM system is that the receiver
employs a simple GD which can perform symbol detection with low complexity
and with a minimal CSI requirement. However, as will be shown later,
if a conventional QAM constellation is used, the system shows a drop
in error rate performance with higher modulation orders, since the
symbols with lowest energy in the QAM constellation dominate the performance
of the GD. This phenomenon has a greater impact in the case of RIS-RQSM
than in the RIS-SM system of \cite{basar2020reconfigurable}, as in
the former a higher average energy is received at the non-selected
antennas, which results in reducing the performance of the GD. This
fact motivates us to design a new QAM constellation in order to favor
the GD\footnote{Both the ML detector and the GD perform better with the proposed constellation,
but the GD benefits more significantly.}. Hence, in this section we optimize the constellation to minimize
the BER of the RIS-RQSM system with GD. In order to lower the complexity,
we employ a number of approximations in this section to simplify the
ABEP upper bound which will then serve as our objective function.
However, the extensive numerical results included in Table~\ref{tab:Comparison}
and in the next section verify the accuracy of these approximations
and show that the proposed approach is practical and yields excellent
results.

\begin{figure}[t]
\begin{centering}
\includegraphics[scale=0.15]{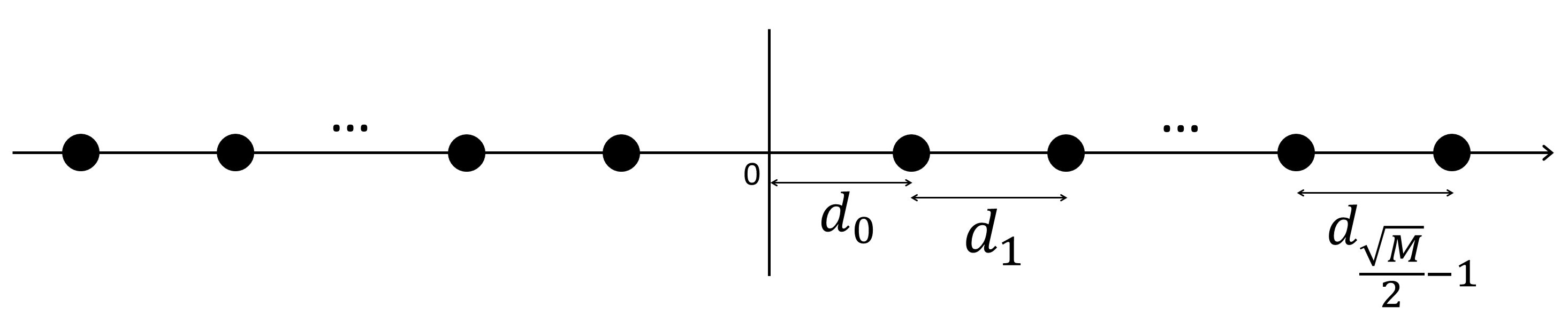}
\par\end{centering}
\caption{Normalized PAM constellation design for the RIS-RQSM system.\label{fig:constellation}}

\end{figure}

Thanks to the symmetry in the RIS-RQSM system, the real and imaginary
dimensions of the constellation can be designed separately following
the same method, which simplifies the optimization procedure. Hence,
the optimization problem is defined as 
\begin{align}
\underset{\mathcal{M}_{R}}{\min} & \ \mathrm{ABEP}_{\mathrm{ub}}\label{eq: min op problem-constellation}\\
\mbox{s.t.}\;\; & \ \sum_{i=1}^{\sqrt{M}}\left(x^{\mathcal{R}}\right)^{2}\leq\frac{\sqrt{M}E_{s}}{2},\nonumber 
\end{align}
where $\mathrm{ABEP}_{\mathrm{ub}}$ is the approximate upper bound
on the ABEP expressed in (\ref{eq:ABEP}). It is trivial to observe
that the signal constellation should be symmetric about the origin.
Therefore, we define the one-dimensional ``normalized'' $\sqrt{M}$-PAM
constellation for the real and imaginary dimensions according to Fig.~\ref{fig:constellation},
such that the minimum-energy symbol has distance $d_{0}\sqrt{E_{s}}$
from the origin, while the distance between the $i$-th and $(i+1)$-th
symbols is denoted by $d_{i}\sqrt{E_{s}}$, $i=1,2,\dots,\frac{\sqrt{M}}{2}-1$.
Due to the symmetry about the origin, there exist $\sqrt{M}/2$ parameters
that need to be optimized. For example, in 2-PAM, there is only one
parameter $d_{0}$; it is clear that in this case $d_{0}=1/\sqrt{2}$,
so that this optimization framework is not necessary in that case.
In a 4-PAM constellation there are two parameters $d_{0}$ and $d_{1}$
that should be optimized such that $d_{0}$ is increased and $d_{1}$
is decreased with respect to the values for conventional PAM, i.e.,
the two ``inner'' symbols are moved further away from the origin
and the two ``outer'' symbols are moved towards the origin; this
adjustment of the constellation points provides a balance between
the spatial domain symbol error probability and the IQ modulation
domain symbol error probability.

The expression for $\mathrm{ABEP}_{\mathrm{ub}}$ in (\ref{eq:ABEP})
is a relatively complex function of the parameters $\left\{ d_{i}\right\} $
due to the summation over all symbols in calculating $P_{e}\left(m\right)$
and in calculating $\mathrm{PEP}\left(x^{\mathcal{R}}\rightarrow\hat{x}^{\mathcal{R}}|m=\hat{m}\right)$
associated with all of these distances. Hence, to simplify the solution
for the optimization problem in (\ref{eq: min op problem-constellation})
we adopt some accurate approximations for evaluating the upper bound
on the ABEP that are valid at high SNR and with large $N$.

It is well-known that at high SNR values the IQ modulation domain
bit error probability (BEP) is dominated by the pairs of constellation
points separated by the minimum Euclidean distance, and it is also
clear that the minimum-energy symbols control the BEP in the spatial
domain. Hence, considering Gray coding for the constellation, an approximate
upper bound on the ABEP is given by 
\begin{equation}
\mathrm{ABEP}_{\mathrm{ub}}\approx\frac{4}{\sqrt{M}\log_{2}\left(\sqrt{M}N_{r}\right)}\sum_{i=1}^{\frac{\sqrt{M}}{2}-1}\mathrm{Q}\left(\sqrt{\frac{\beta^{2}E_{s}d_{i}^{2}}{2N_{0}}}\right)+0.5\tilde{P}_{e}\left(m\right),\label{eq:Approx_ABEP}
\end{equation}
where $\tilde{P}_{e}\left(m\right)$ is the corresponding approximate
value of $P_{e}\left(m\right)$, given by 
\begin{equation}
\tilde{P}_{e}\left(m\right)=\frac{4\left(N_{r}-1\right)}{M}\sum_{\delta\in\mathcal{D}_{d_{0}\sqrt{E_{s}}}}\mathrm{PEP}\left(m\rightarrow\hat{m}|x^{\mathcal{R}}=d_{0}\sqrt{E_{s}},\delta\right),\label{eq:APPROX PEP mm_hat}
\end{equation}
where $\mathcal{D}_{d_{0}\sqrt{E_{s}}}=\Bigl\{1,\frac{d_{0}}{d_{0}+d_{1}},\dots,\frac{d_{0}}{d_{0}+d_{1}+\cdots+d_{\frac{\sqrt{M}}{2}-1}}\Bigr\}$,
and we use the fact that $\tilde{P}_{e}\left(m\right)\ll1$, hence
$1-\tilde{P}_{e}\left(m\right)\approx1$ (note that the optimization
function increases the distance between two inner symbols, so that
in (\ref{eq:Approx_ABEP}), we did not consider the distance between
the pair of inner symbols as the minimum distance). Then, the optimization
problem can be updated as 
\begin{align}
\underset{\left\{ d_{i}\right\} }{\min} & \ \mathrm{ABEP}_{\mathrm{ub}}\mbox{ in (\ref{eq:Approx_ABEP})}\label{eq: min op problem-constellation-app}\\
\mbox{s.t.}\;\; & \ \sum_{i=0}^{\frac{\sqrt{M}}{2}-1}\left(\sum_{j=0}^{i}d_{j}\right)^{2}\leq\frac{\sqrt{M}}{4}.\nonumber 
\end{align}

Solving the above optimization problem is not a straightforward task
and requires the use of exhaustive search methods. However, standard
lattice constellation structures, such as QAM or PAM, suggest that
equal distances between adjacent pairs of symbols admit a very simple
approach which provides a near-optimal solution in terms of the symbol
error rate performance. Hence, in the following, we assume that the
distances between ``positive'' adjacent symbols are equal (it is
worth recalling that there is a symmetry about the origin, hence the
distances between negative adjacent symbols are also equal).

\subsection*{Special case where $d_{1}=d_{2}=\dots=d_{\frac{\sqrt{M}}{2}-1}$}

In this case, the problem consists of optimizing the two variables
$d_{0}$ and $d_{1}$. Hence, the optimization problem reduces to
\begin{align}
\underset{\left\{ d_{0},d_{1}\right\} }{\min} & \ \frac{4}{\sqrt{M}\log_{2}\left(\sqrt{M}N_{r}\right)}M'\mathrm{Q}\left(\sqrt{\frac{\beta^{2}E_{s}d_{1}^{2}}{2N_{0}}}\right)+0.5\tilde{P}_{e}\left(m\right),\label{eq:op. problem-d0 d1}\\
\mbox{s.t.}\;\; & \ 2d_{0}^{2}+\frac{M'\left(2M'+1\right)}{3}d_{1}^{2}+2M'd_{0}d_{1}\leq1,\nonumber 
\end{align}
where we define $M'=\frac{\sqrt{M}}{2}-1.$ From the inequality constraint,
$d_{1}$ can be obtained as a function of $d_{0}$ (here we force
equality in the constraint above to maximize the achievable SNR at
the receiver. It will be shown later in this section that equality
indeed holds at the optimum point). Then, by performing a grid search
over variable $d_{0}$, we can find the minimum value of the $\mathrm{ABEP}_{\mathrm{ub}}$.
However, taking the equal positive distance into account, it is more
valuable to find an analytical solution; this is the subject of the
remainder of this section.

\subsubsection*{Analytical approach - asymptotic analysis}

In order to find an efficient analytical solution for the optimization
problem, we analyze the distributions of $Z_{1}$ and $Z_{2}$ in
more detail in order to obtain a more tractable approximate expression
for $\mathrm{ABEP}_{\mathrm{ub}}$. We see that the variance of $Z_{2}$
(i.e., the average received energy of the signal on a non-selected
receive antenna) in the event $\mathcal{E}_{1}$ increases with decreasing
$\bar{\lambda}=\frac{\delta^{2}}{1+\delta^{2}}$, or equivalently,
with decreasing $\delta=\left|\frac{x^{\mathcal{R}}}{x^{\mathcal{I}}}\right|$;
in other words, $\rho_{1}^{2}$ in (\ref{eq:rho_1^2}) is maximized
when $\delta$ is minimized. There are two consequences of this fact:
first, the BEP related to the spatial domain is dominated by those
symbols bearing the minimum energy in the real part while their corresponding
imaginary parts have the maximum energy, i.e., the PEP associated
with $\delta_{\min}=\frac{\min\left|x^{\mathcal{R}}\right|}{\max\left|x^{\mathcal{I}}\right|}=\frac{d_{0}}{d_{0}+M'd_{1}}$
dominates (\ref{eq:APPROX PEP mm_hat}); secondly, comparing the two
events $\mathcal{E}_{1}$ and $\mathcal{E}_{2}$, the event $\mathcal{E}_{2}$
has a minor impact on the value of $\mathrm{PEP}\left(m\rightarrow\hat{m}|x^{\mathcal{R}},\delta\right)$,
as the variance of $Z_{2}$ in the event $\mathcal{E}_{1}$ is significantly
greater than that in the event $\mathcal{E}_{2}$ due to the appearance
of $\bar{\lambda}$ in the denominator. In summary, considering the
above comments, the PEP associated to $\delta_{\min}$ dominates and
the event $\mathcal{E}_{2}$ can be eliminated from the PEP analysis,
therefore $\tilde{P}_{e}\left(m\right)$ can be approximated as 
\begin{align}
\tilde{P}_{e}\left(m\right) & \approx\frac{4\left(N_{r}-1\right)}{M}\mathrm{PEP}\left(m\rightarrow\hat{m}|x^{\mathcal{R}}=d_{0}\sqrt{E_{s}},\delta=\delta_{\min}\right)\nonumber \\
 & \approx\frac{4\left(N_{r}-1\right)^{2}}{MN_{r}}\mathrm{I}_{1}\left(\mu_{1},\rho_{1},\sigma_{1}\right).\label{eq:approx-P_e(m)}
\end{align}
In addition, by substituting $\bar{\lambda}=\frac{\delta_{\min}^{2}}{1+\delta_{\min}^{2}}$
into (\ref{eq:rho_1^2}) and performing some minor algebraic manipulations,
the variance of $Z_{2}$ in the event $\mathcal{E}_{1}$ can be expressed
as 
\[
\rho_{1}^{2}=\frac{NE_{s}}{2}\left(d_{0}^{2}+\left(d_{0}+M'd_{1}\right)^{2}\right)+\frac{N_{0}}{2}.
\]
Note that $\bar{E}\triangleq d_{0}^{2}+\left(d_{0}+M'd_{1}\right)^{2}$
is the sum of the energies associated with the symbols with minimum
and maximum distance from the origin. It is clear that $1\leq\bar{E}<\epsilon_{M}$,
where equality holds for $M=16$, and $\epsilon_{M}$ is defined as
the total energy of the inner and outer symbols in the conventional
$\sqrt{M}$-PAM constellation (since the conventional constellation
is the worst-case scenario, $\bar{E}$ can be upper bounded by $\epsilon_{M}$),
so that we obtain $\epsilon_{M}=\frac{3(M-2\sqrt{M}+2)}{2(M-1)}$
(note that the average energy of the PAM constellation is $1/2$).
Therefore, we can write 
\[
\frac{NE_{s}}{2}+\frac{N_{0}}{2}\leq\rho_{1}^{2}<\frac{NE_{s}}{2}\epsilon_{M}+\frac{N_{0}}{2}.
\]
In addition, it is easy to prove that (\ref{eq:approx-P_e(m)}) is
monotonically increasing with respect to $\rho_{1}$. Hence, $\tilde{P}_{e}\left(m\right)$
can be expressed as 
\begin{align*}
\tilde{P}_{e}\left(m\right) & \approx\left.\tilde{P}_{e}\left(m\right)\right|_{\rho_{1}^{2}=\frac{NE_{s}}{2}+\frac{N_{0}}{2}},\ M=16,\\
\tilde{P}_{e}\left(m\right) & \lesssim\left.\tilde{P}_{e}\left(m\right)\right|_{\rho_{1}^{2}=\frac{NE_{s}}{2}\epsilon_{M}+\frac{N_{0}}{2}},\ M>16.
\end{align*}

Finally, from the formula $\sigma_{1}^{2}=N\left(x^{\mathcal{R}}\right)^{2}\frac{4-\pi}{4}+\frac{N_{0}}{2}$
applied to the minimum energy symbol $x^{\mathcal{R}}=d_{0}\sqrt{E_{s}}$
and considering the fact that $d_{0}^{2}\ll1$, the variance of $Z_{1}$
can be approximated as $\sigma_{1}^{2}\approx\frac{N_{0}}{2}$\footnote{Here we are assuming that $N$ is sufficiently large so the SNR range
$\frac{\frac{4-\pi}{2}NE_{s}d_{0}^{2}}{N_{0}}\ll1$ is of interest,
i.e., the BER is extremely low outside of this SNR range.}.

Therefore, after some manipulations we obtain $\tilde{P}_{e}\left(m\right)$
as 

\begin{align*}
\tilde{P}_{e}\left(m\right)\approx & \frac{2\left(N_{r}-1\right)^{2}}{MN_{r}}\left(\frac{1}{3}\sqrt{\frac{NE_{s}+N_{0}}{NE_{s}+2N_{0}}}e^{-\frac{\pi N^{2}E_{s}d_{0}^{2}}{4NE_{s}+8N_{0}}}+\sqrt{\frac{NE_{s}+N_{0}}{NE_{s}+\frac{7}{3}N_{0}}}e^{-\frac{\pi N^{2}E_{s}d_{0}^{2}}{3NE_{s}+7N_{0}}}\right),\ M=16,\\
\tilde{P}_{e}\left(m\right)\lesssim & \frac{2\left(N_{r}-1\right)^{2}}{MN_{r}}\left(\frac{1}{3}\sqrt{\frac{NE_{s}\epsilon_{M}+N_{0}}{NE_{s}\epsilon_{M}+2N_{0}}}e^{-\frac{\pi N^{2}E_{s}d_{0}^{2}}{4NE_{s}\epsilon_{M}+8N_{0}}}+\sqrt{\frac{NE_{s}\epsilon_{M}+N_{0}}{NE_{s}\epsilon_{M}+\frac{7}{3}N_{0}}}e^{-\frac{\pi N^{2}E_{s}d_{0}^{2}}{3NE_{s}\epsilon_{M}+7N_{0}}}\right),\\
 & M>16.
\end{align*}
Also applying the exponential approximation of the Q-function in (\ref{eq:op. problem-d0 d1}),
the optimization problem becomes 
\begin{align}
\underset{\left\{ d_{0},d_{1}\right\} }{\min} & \ \mathrm{ABEP}_{\mathrm{ub}}\approx a_{0}e^{-b_{0}d_{0}^{2}}+a_{1}e^{-b_{1}d_{0}^{2}}+M'a_{2}\left(\frac{1}{12}e^{-b_{2}d_{1}^{2}}+\frac{1}{4}e^{-\frac{4}{3}b_{2}d_{1}^{2}}\right)\label{eq:min problem M=00003D16}\\
\mbox{s.t.}\;\; & \ 2d_{0}^{2}+\frac{M'\left(2M'+1\right)}{3}d_{1}^{2}+2M'd_{0}d_{1}\leq1,\nonumber 
\end{align}
where we define 
\begin{align*}
 & a_{0}=\frac{\left(N_{r}-1\right)^{2}}{3MN_{r}}\sqrt{\frac{NE_{s}\epsilon_{M}+N_{0}}{NE_{s}\epsilon_{M}+2N_{0}}},\;b_{0}=\frac{\pi N^{2}E_{s}}{4NE_{s}\epsilon_{M}+8N_{0}},\\
 & a_{1}=\frac{\left(N_{r}-1\right)^{2}}{MN_{r}}\sqrt{\frac{NE_{s}\epsilon_{M}+N_{0}}{NE_{s}\epsilon_{M}+\frac{7}{3}N_{0}}},\;b_{1}=\frac{\pi N^{2}E_{s}}{3NE_{s}\epsilon_{M}+7N_{0}},\\
 & a_{2}=\frac{4}{\sqrt{M}\log_{2}\left(\sqrt{M}N_{r}\right)},\ b_{2}=\frac{\pi N^{2}E_{s}}{16N_{0}}.
\end{align*}

The problem in (\ref{eq:min problem M=00003D16}) is not a convex
optimization problem, as the objective function is not convex in the
domain of $d_{0},d_{1}\in\mathbb{R}_{+}$. However, it is easy to
see that (\ref{eq:min problem M=00003D16}) satisfies the convexity
condition $\nabla^{2}\mathrm{ABEP_{ub}}\geq0$ when $d_{0}\geq\frac{1}{\sqrt{2b_{i}}}$,
$i=0,1$, and $d_{1}\geq\frac{1}{\sqrt{2b_{2}}}$. For sufficiently
high values of $\frac{N^{2}E_{s}}{N_{0}}$ (note that $N\gg1$), it
can be concluded that $\left\{ b_{0},b_{1},b_{2}\right\} $ are sufficiently
large such that the optimized $\left\{ d_{0},d_{1}\right\} $ lie
in the convex region of the objective function. For such $\left\{ b_{0},b_{1},b_{2}\right\} $,
the problem is convex and can be solved using the following procedure.

The \ac{KKT} \cite{boyd2004convex} conditions associated to the
above problem hold and are given by 
\begin{align*}
 & 1.\;f_{1}\left(d_{0}^{\star},d_{1}^{\star}\right)\leq0;\quad2.\;\nu^{\star}\geq0;\quad3.\;\nu^{\star}f_{1}\left(d_{0}^{\star},d_{1}^{\star}\right)=0;\\
 & 4.\;-2a_{0}b_{0}d_{0}^{\star}e^{-b_{0}d_{0}^{\star^{2}}}-2a_{1}b_{1}d_{0}^{\star}e^{-b_{1}d_{0}^{\star^{2}}}+\nu^{\star}\left(4d_{0}^{\star}+2M'd_{1}^{\star}\right)=0;\\
 & 5.\;-\frac{1}{6}M'a_{2}b_{2}d_{1}^{\star}e^{-b_{2}d_{1}^{\star^{2}}}-\frac{2}{3}M'a_{2}b_{2}d_{1}^{\star}e^{-\frac{4}{3}b_{2}d_{1}^{\star^{2}}}+\nu^{\star}\left(\frac{2M'\left(2M'+1\right)}{3}d_{1}^{\star}+2M'd_{0}^{\star}\right)=0;
\end{align*}
where $\nu$ is the Lagrange multiplier associated with the inequality
constraint. From condition 3, we see that $\nu^{\star}=0$ or $f_{1}\left(d_{0}^{\star},d_{1}^{\star}\right)=0$.
However, if $\nu^{\star}=0$, from conditions 4 and 5 we obtain $d_{0}^{\star}=d_{1}^{\star}=+\infty$,
where clearly contradicts condition 1. Therefore, we have 
\[
2d_{0}^{\star^{2}}+\frac{M'\left(2M'+1\right)}{3}d_{1}^{\star^{2}}+2M'd_{0}^{\star}d_{1}^{\star}-1=0,
\]
which yields 
\begin{equation}
d_{0}^{\star}=\frac{-2M'd_{1}^{\star}+\sqrt{4M^{\prime^{2}}d_{1}^{\star^{2}}-8\left(\frac{M'\left(2M'+1\right)}{3}d_{1}^{\star^{2}}-1\right)}}{4}.\label{eq:d0}
\end{equation}
Then, from conditions 4 and 5, we obtain 
\begin{equation}
\nu^{\star}=a_{0}b_{0}d_{0}^{\star^{2}}e^{-b_{0}d_{0}^{\star^{2}}}+a_{1}b_{1}d_{0}^{\star^{2}}e^{-b_{1}d_{0}^{\star^{2}}}+\frac{1}{12}M'a_{2}b_{2}d_{1}^{\star^{2}}e^{-b_{2}d_{1}^{\star^{2}}}+\frac{1}{3}M'a_{2}b_{2}d_{1}^{\star^{2}}e^{-\frac{4}{3}b_{2}d_{1}^{\star^{2}}}.\label{eq:nu}
\end{equation}
Substituting for $\nu^{\star}$ from (\ref{eq:nu}) and subsequently
for $d_{0}^{\star}$ from (\ref{eq:d0}) into condition 5, the optimization
problem reduces to a single-variable equation in $d_{1}^{\star}$.
This equation does not admit a closed-form analytical solution; however
it is easy to solve numerically.

We conclude this section by providing a numerical example in Table~\ref{tab:Comparison}.
In this table, we compare the optimal $\{d_{i}\}$ obtained by an
exhaustive search to minimize the ABEP in (\ref{eq:ABEP}) with the
corresponding values with equal positive distances obtained via the
proposed analytical approach, where $N=256$, $N_{r}=4$ and $M=64$.
It can be seen that positive distances $\{d_{i}\}$, $i>0$, obtained
via exhaustive search are almost equal, and that these values become
more similar with increasing SNR. In addition, the ABEP values acquired
by using the optimal values from the proposed analytical approach
are quite comparable to the equivalent ABEP obtained by optimal values
of the grid search, which serves as a proof that the assumptions we
made to offer a straightforward analytical solution to the optimization
problem were indeed accurate.

\begin{table*}[t]
\caption{Comparison between optimal $\{d_{i}\}$ values obtained via minimizing
(\ref{eq:ABEP}) by grid search and the corresponding values obtained
by the analytical approach of (\ref{eq:min problem M=00003D16}),
where $N=256$, $N_{r}=4$ and $M=64$.\label{tab:Comparison}}

\centering{}%
\begin{tabular}{|c|c|c|c|c|c|>{\centering}p{2cm}|>{\centering}p{2cm}|c|}
\hline 
\multicolumn{6}{|c|}{Minimized ABEP based on (\ref{eq:ABEP}) using grid search} &
\multicolumn{3}{c|}{Minimized ABEP by using analytical approach of (\ref{eq:min problem M=00003D16})}\tabularnewline
\hline 
\hline 
SNR (dB) &
$d_{0}$ &
$d_{1}$ &
$d_{2}$ &
$d_{3}$ &
ABEP &
$d_{0}$ &
$d_{1}$ &
ABEP\tabularnewline
\hline 
-23 &
0.2609 &
0.250 &
0.257 &
0.272 &
$6.97\times10^{-4}$ &
0.2481 &
0.2632 &
$7.62\times10^{-4}$\tabularnewline
\hline 
-21 &
0.2695 &
0.248 &
0.253 &
0.262 &
$6.46\times10^{-5}$ &
0.2661 &
0.2543 &
$6.67\times10^{-5}$\tabularnewline
\hline 
-19 &
0.2890 &
0.240 &
0.243 &
0.248 &
$2.90\times10^{-6}$ &
0.2891 &
0.2426 &
$2.96\times10^{-6}$\tabularnewline
\hline 
-17 &
0.3169 &
0.227 &
0.228 &
0.232 &
$5.71\times10^{-8}$ &
0.3179 &
0.2278 &
$5.82\times10^{-8}$\tabularnewline
\hline 
\end{tabular}
\end{table*}

\section{Numerical Results\label{sec:Numerical-Results}}

In this section, we demonstrate the error rate performance of the
proposed RIS-RQSM system via numerical simulations. First, we investigate
the performance of the proposed RIS-RQSM system using conventional
QAM constellations and provide comparisons with corresponding systems
using QAM constellations that are optimized based on the approach
proposed in Section~\ref{sec:IQ-Modulation-Design}. Next, we compare
the results obtained by the optimized constellations with the error
rate performance of the most prominent recently proposed RIS-SM \cite{basar2020reconfigurable}
system, which serves as the benchmark scheme for the proposed approach.

\begin{figure}[t]
\begin{centering}
\includegraphics[scale=0.6]{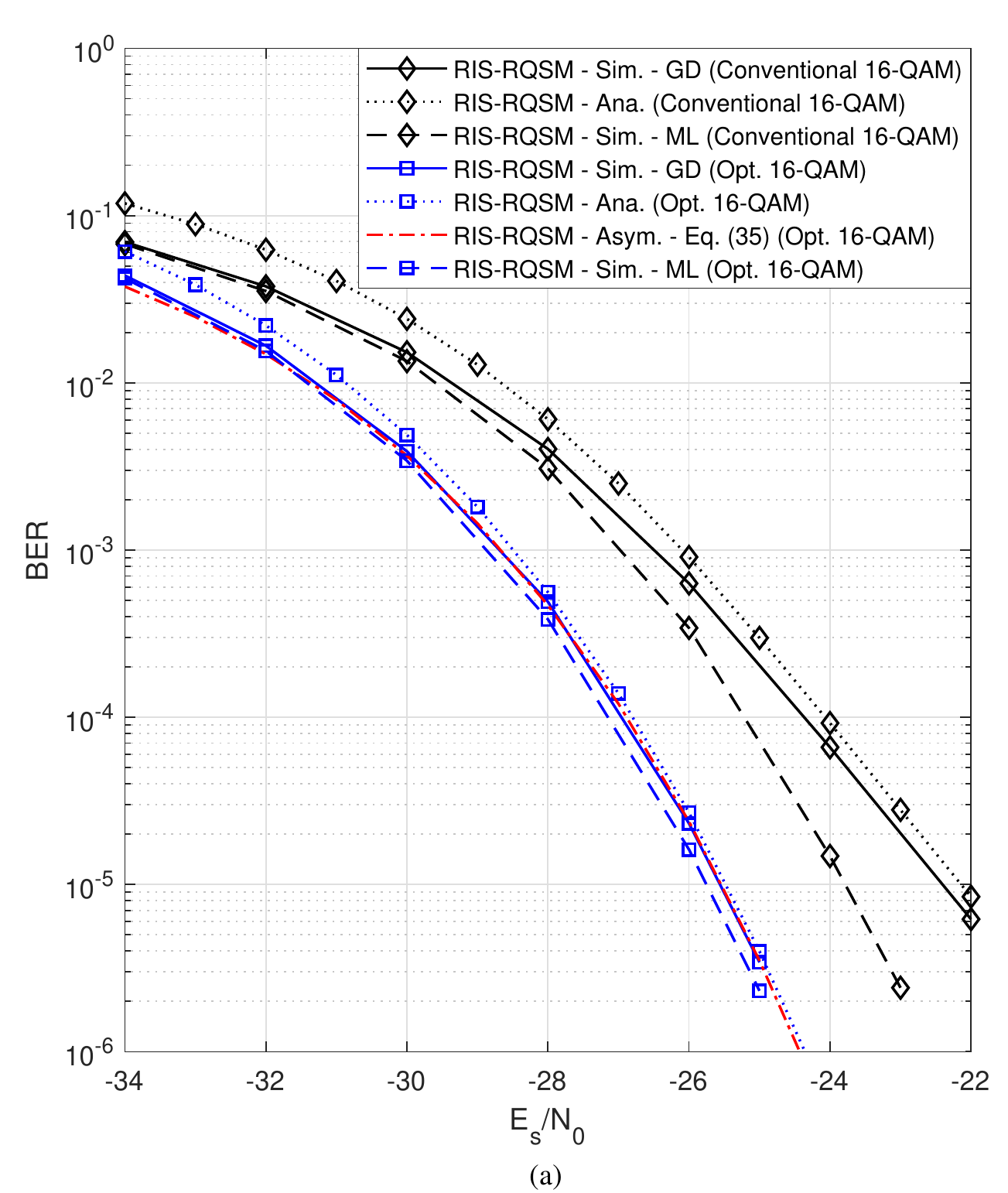}\includegraphics[scale=0.6]{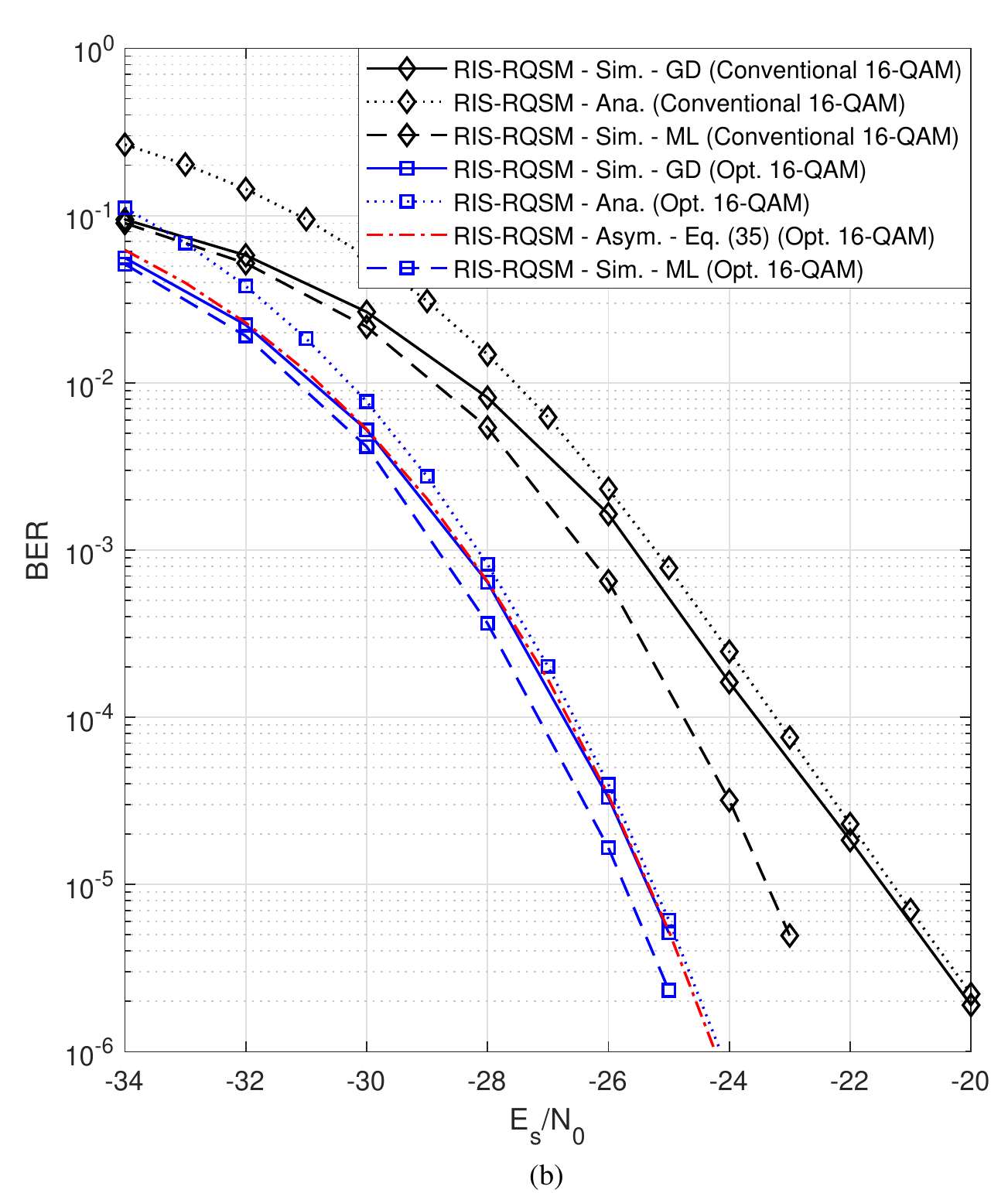}
\par\end{centering}
\caption{Analytical and simulation BER results of the proposed RIS-RQSM system
with and without optimized constellation. Here $M=16$, $N=256$,
and (a) $N_{r}=4$ ($R=8$~bpcu), (b) $N_{r}=8$ ($R=10$~bpcu).\label{fig:Analytical-and-simulation-16}}
\end{figure}
\begin{figure}[t]
\begin{centering}
\includegraphics[scale=0.6]{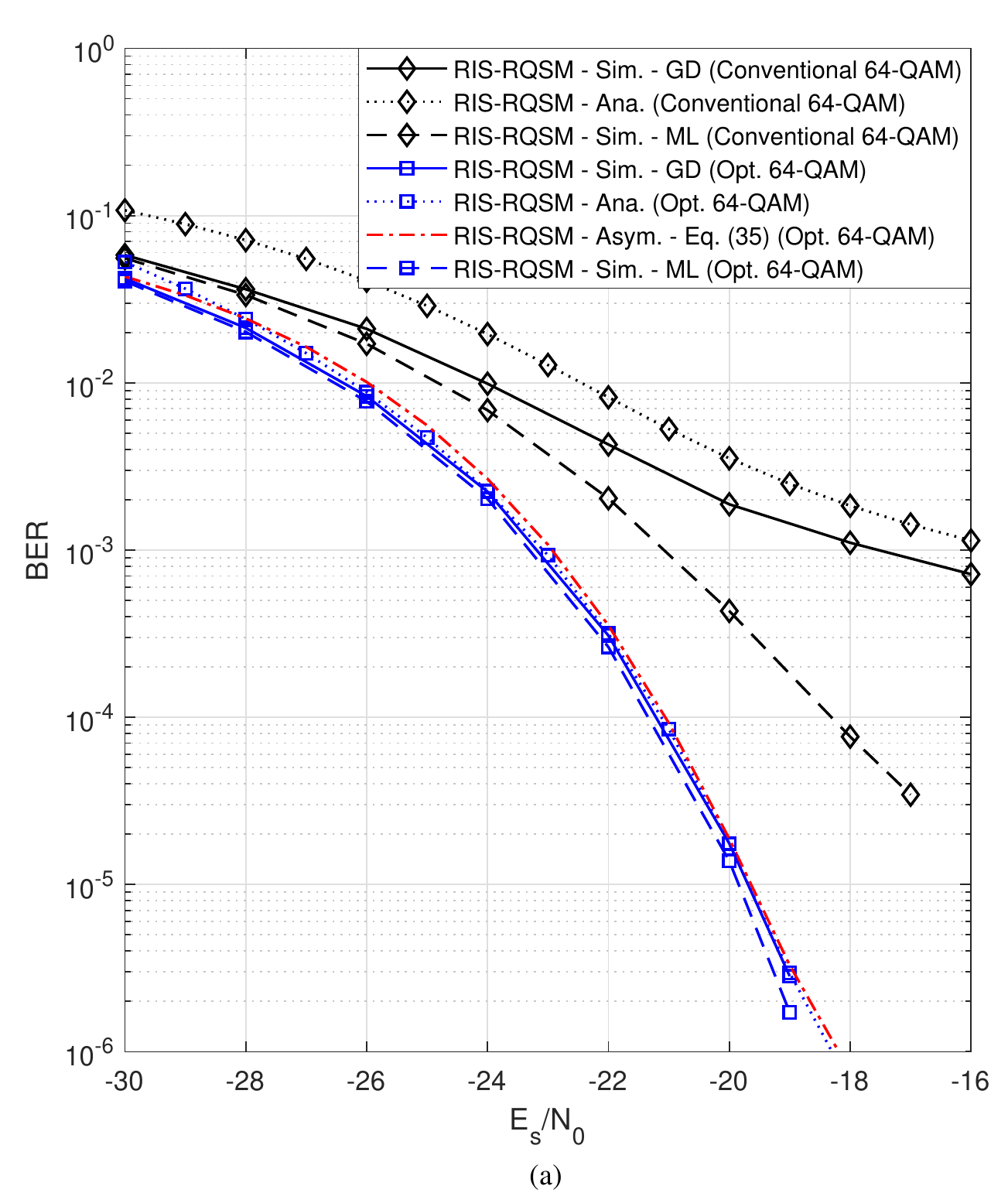}\includegraphics[scale=0.6]{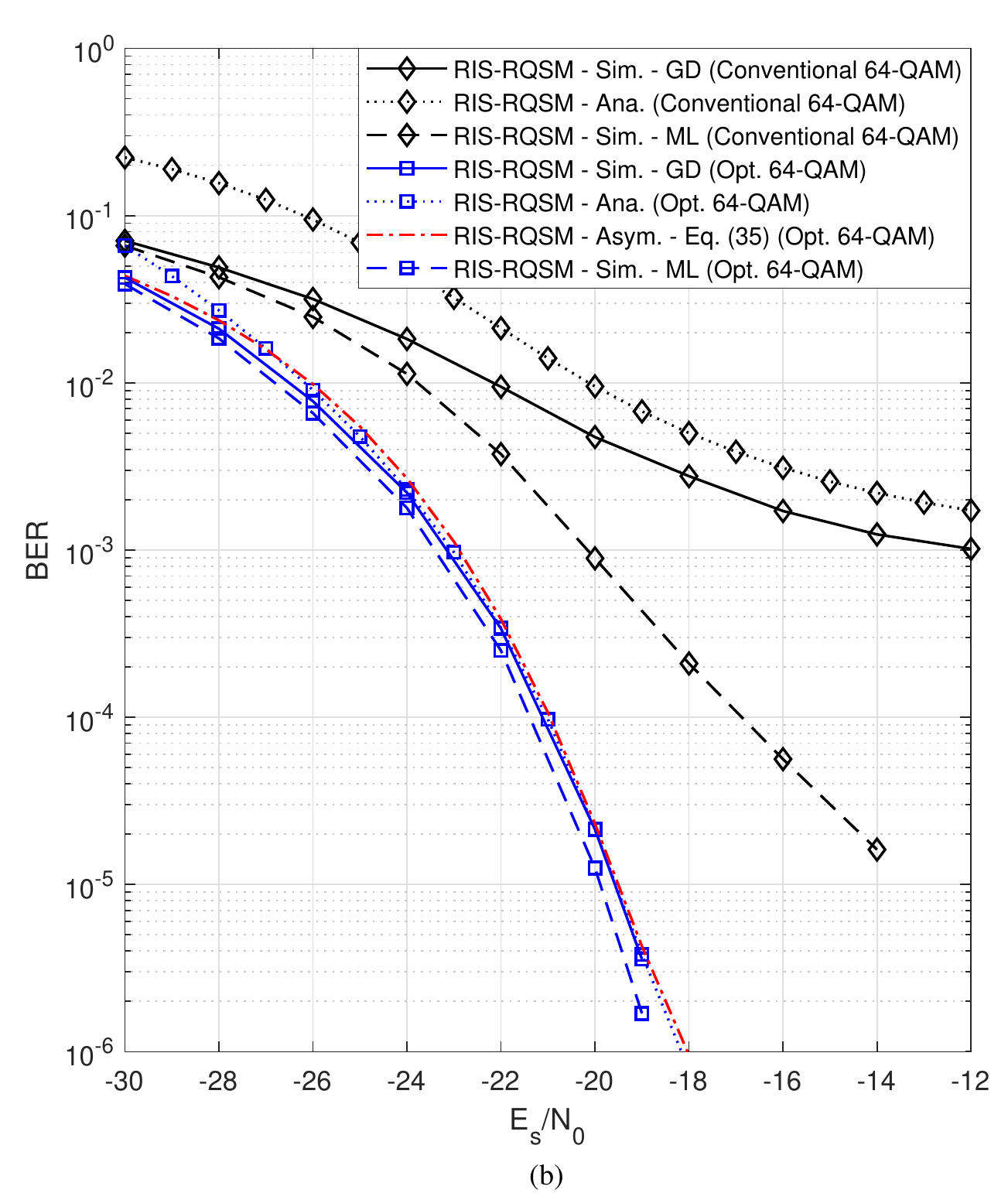}
\par\end{centering}
\caption{Analytical and simulation BER results of the proposed RIS-RQSM system
with and without optimized constellation. Here $M=64$, $N=256$,
and (a) $N_{r}=4$ ($R=10$~bpcu), (b) $N_{r}=8$ ($R=12$~bpcu).\label{fig:Analytical-and-simulation-64}}
\end{figure}

Fig.~\ref{fig:Analytical-and-simulation-16} shows the BER performance
of the proposed RIS-RQSM system with $N=256$ for the cases of $N_{r}=4$
and $N_{r}=8$. In this figure, we also compare the performance of
the RIS-RQSM system using conventional 16-QAM modulation with that
of the system implementing our optimized 16-QAM constellation. The
curves demonstrate the effectiveness of the proposed constellation
design method; it can be observed that optimizing the design of the
constellation significantly enhances the performance of the system.
The proposed constellation for RIS-RQSM provides approximately 3.2~dB
and 3.8~dB improvement over the conventional constellation in systems
with $N_{r}=4$ and $N_{r}=8$, respectively, at a BER of $10^{-5}$.
We also compare the performance of the GD with that of the ML detector.
We see that there is a very large gap between the performance of the
GD and ML detector in the case of the conventional constellation,
while the performance of the GD in the system using the optimized
constellation is considerably close to that of the ML detector such
that the performance gap is negligible. In order to observe the effect
of optimizing the constellation in a system with higher-order modulation,
we present the BER performance of the RIS-RQSM system with 64-QAM
in Fig.~\ref{fig:Analytical-and-simulation-64}. Here, we see that
in systems with regular QAM constellations, an error floor occurs
with the GD. This is due to the fact that with critical symbols, i.e.,
minimum-energy symbols, $\bar{\lambda}$ can attain a very small value;
hence, non-selected antennas can have a relatively high average received
energy compared to the selected antenna. However, we see that optimizing
the constellation eliminates this error floor and substantially improves
the error rate performance. Similar to systems with 16-QAM constellation,
the performance of the GD is very close to that of ML detector with
optimized constellations. In fact, here the GD becomes feasible only
with the optimized 64-QAM. In Figs.~\ref{fig:Analytical-and-simulation-16} and \ref{fig:Analytical-and-simulation-64},
we also present the analytical ABEP performance of each system. For
systems with conventional QAM constellations, we evaluate and plot
the analytical ABEP upper bounds based on (\ref{eq:ABEP}); we see
that upper bound curves are quite tight and validate the accuracy
of the analytical results. For systems with optimized constellation
we also plot the asymptotic result in (\ref{eq:min problem M=00003D16}).
These curves show that the utilized approximations in Section~\ref{sec:IQ-Modulation-Design}
are completely valid and accurate, especially at high SNR.

\begin{figure}[t]
\begin{centering}
\includegraphics[scale=0.6]{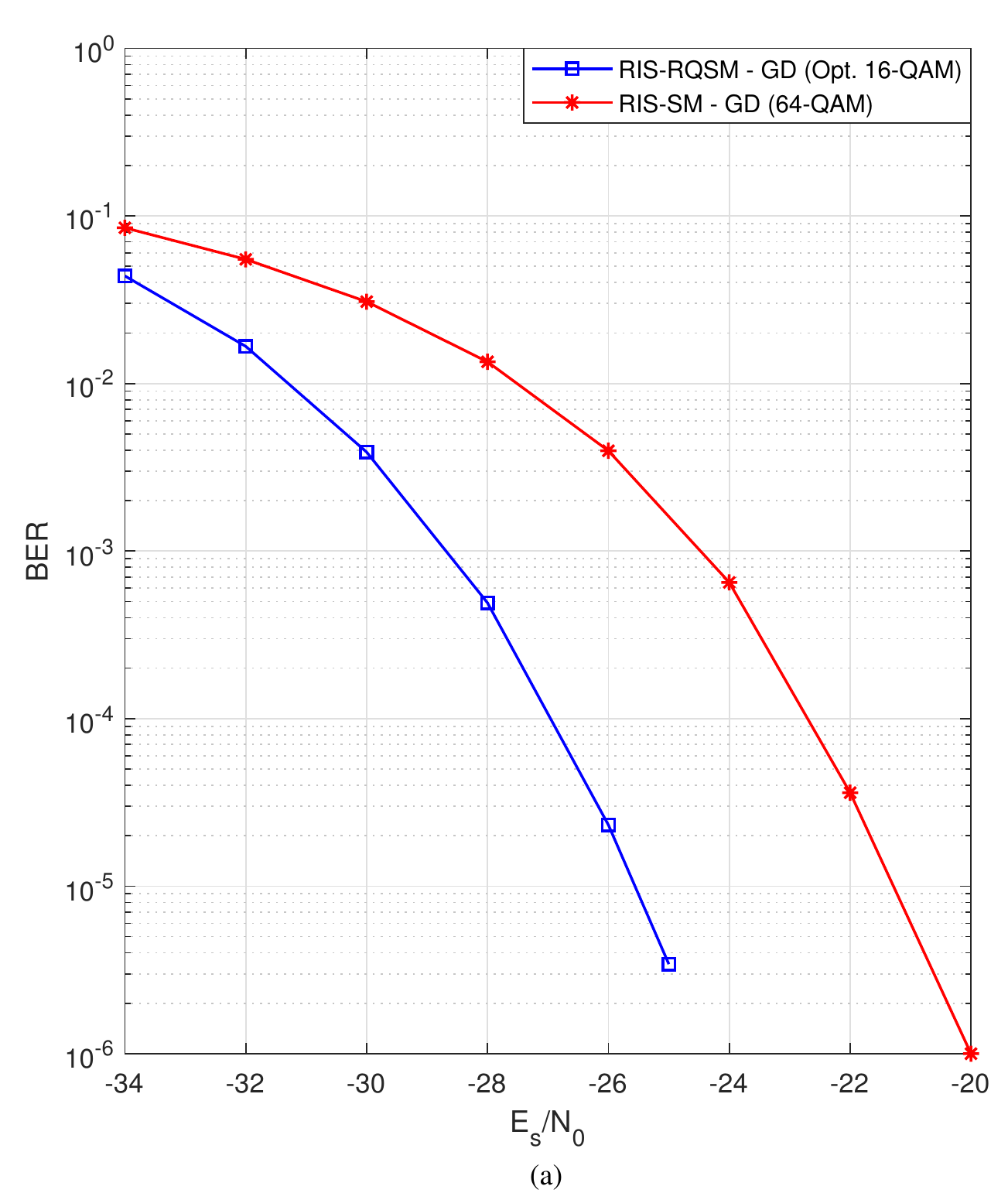}\includegraphics[scale=0.6]{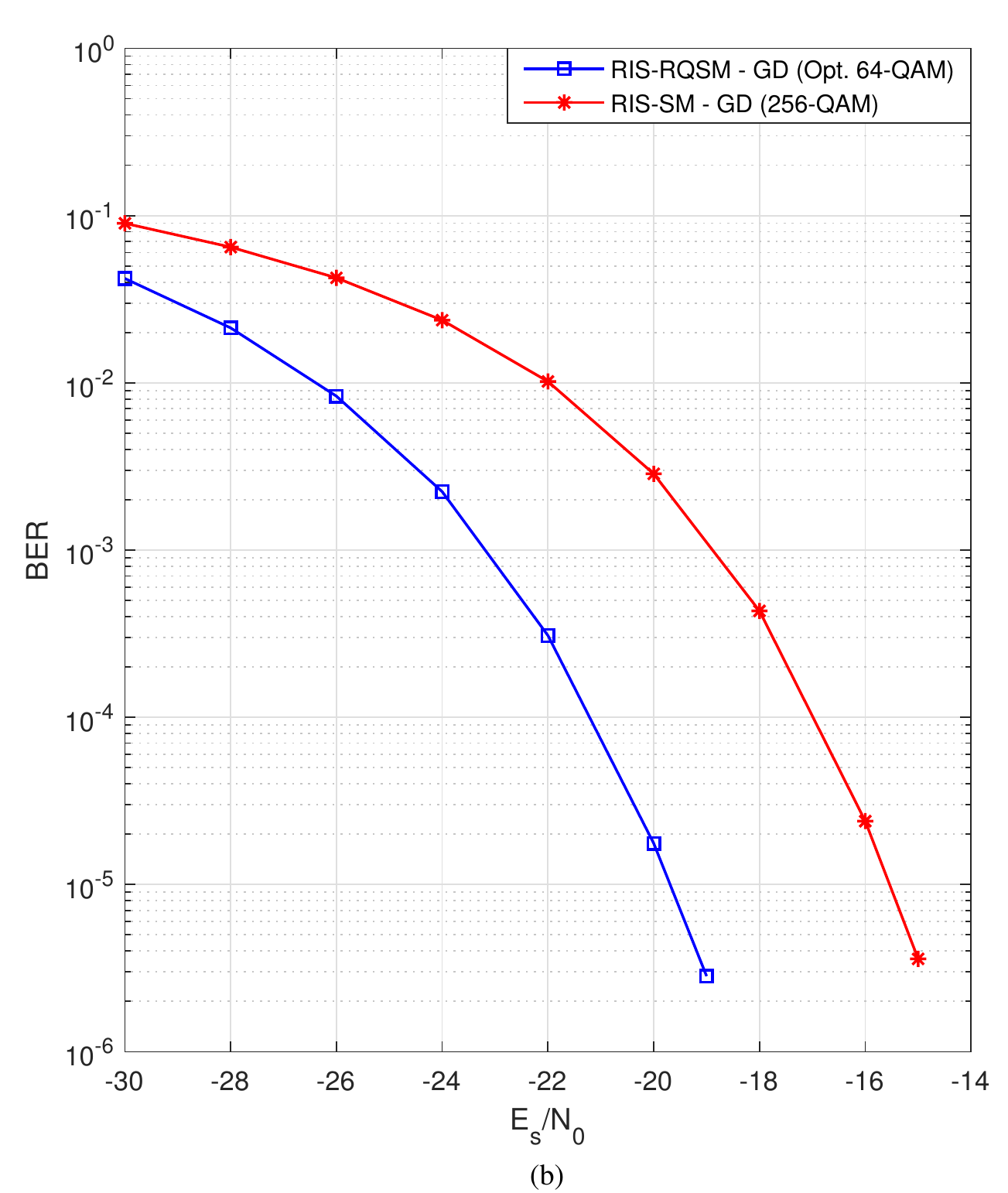}
\par\end{centering}
\caption{Comparison of the BER performance of the proposed RIS-RQSM system
with that of RIS-SM system for $N=256$, $N_{r}=4$, and (a) $R=8$~bpcu,
(b) $R=10$~bpcu.\label{fig:Comparison-Nr4}}
\end{figure}
\begin{figure}[t]
\begin{centering}
\includegraphics[scale=0.6]{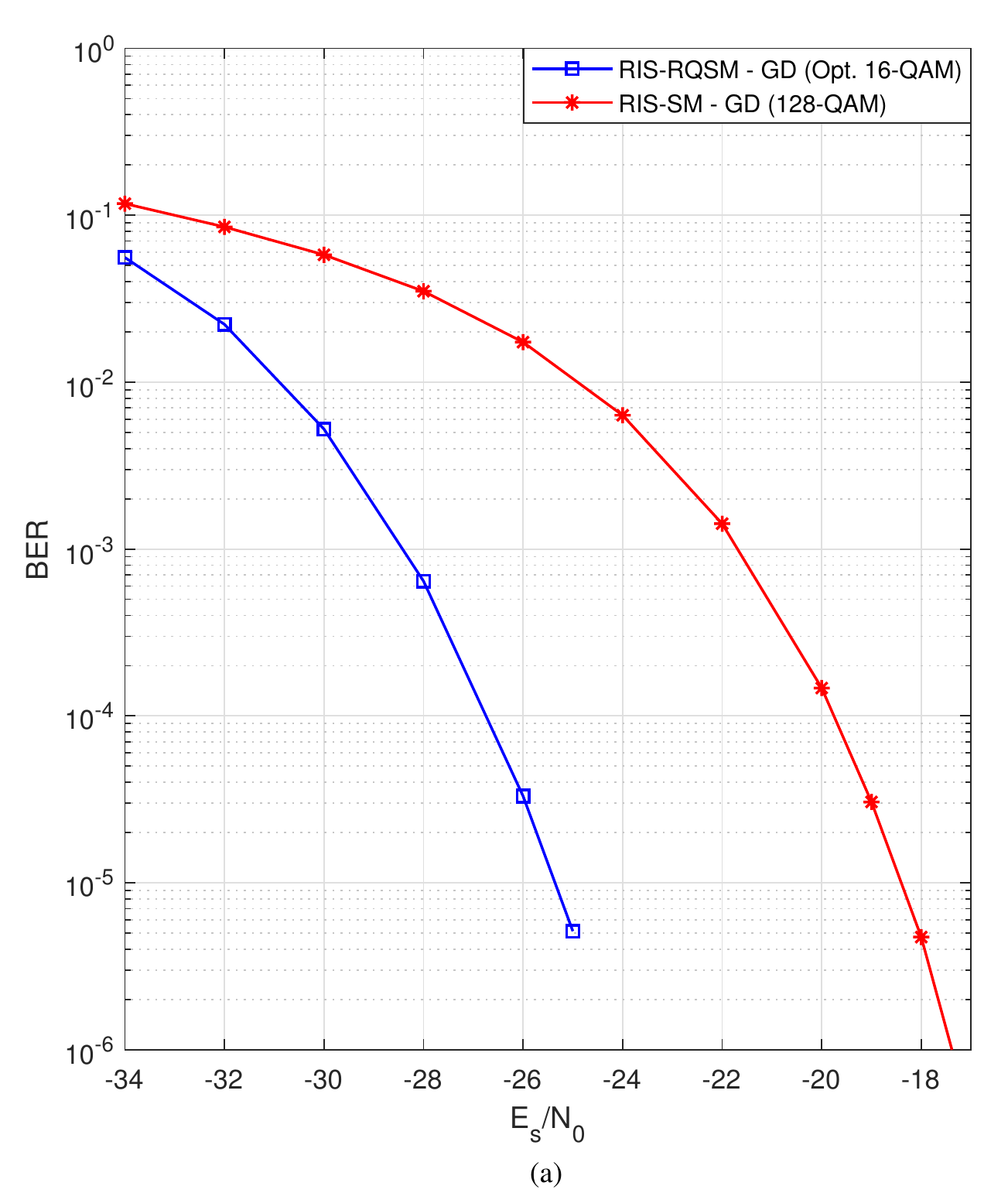}\includegraphics[scale=0.6]{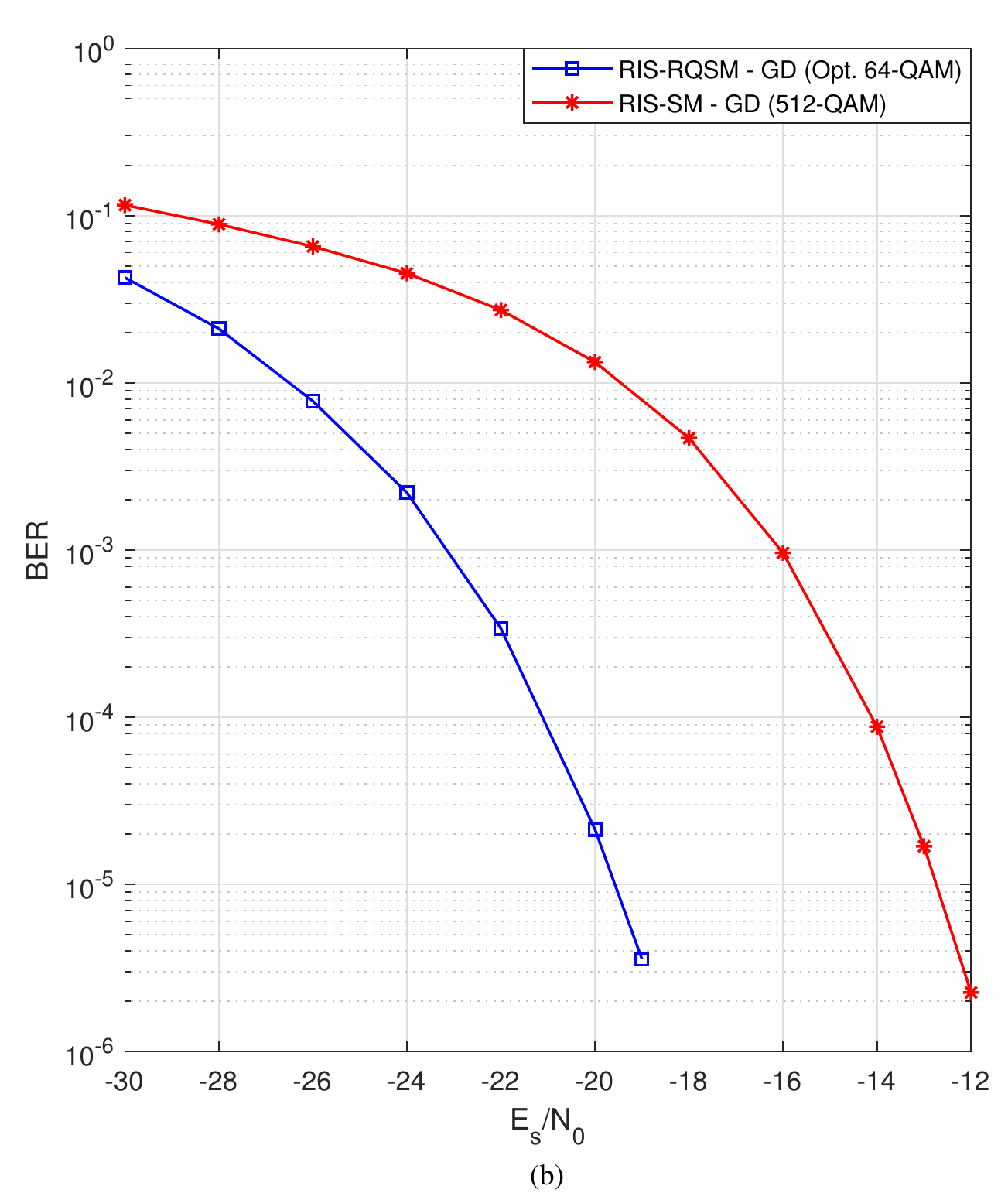}
\par\end{centering}
\caption{Comparison of the BER performance of the proposed RIS-RQSM system
with that of RIS-SM system for $N=256$, $N_{r}=8$, and (a) $R=10$~bpcu,
(b) $R=12$~bpcu.\label{fig:Comparison-Nr8}}
\end{figure}

Next, in Fig.~\ref{fig:Comparison-Nr4}, we compare the BER performance
of the proposed RIS-RQSM system with that of the benchmark scheme,
i.e., RIS-SM, in systems with $N=256$ and $N_{r}=4$. Fig.~\ref{fig:Comparison-Nr4}(a)
shows the performance of the RIS-RQSM and RIS-SM systems where the
bit rate is $R=8$~\ac{bpcu}. Hence, the proposed RIS-RQSM system
uses 16-QAM modulation, while the RIS-SM system uses 64-QAM modulation.
The constellation used in the proposed RIS-RQSM system is optimized
to achieve the best performance. Fig.~\ref{fig:Comparison-Nr4}(b)
presents the performance results in systems with $R=10$~\ac{bpcu},
i.e., where RIS-RQSM and RIS-SM apply 64-QAM and 256-QAM, respectively.
The results show that the proposed RIS-RQSM system substantially outperforms
the benchmark scheme. This is mainly due to the fact that the RIS-SM
system needs to employ a higher-order modulation technique in order
to compensate the additional bits transmitted by the quadrature index
modulation in the proposed RIS-RQSM system. Hence, the superiority
over the benchmark scheme increases by increasing number of receive
antennas, as shown in Fig.~\ref{fig:Comparison-Nr8}. In this figure,
we provide comparisons between the BER performance of the RIS-RQSM
and RIS-SM systems where $N=256$ and $N_{r}=8$. As expected, the
superiority over the RIS-SM system considerably increases in a system
with larger number of receive antennas, as a higher modulation order
is required for the RIS-SM system. The proposed RIS-RQSM system achieves
approximately 4.3~dB and 7~dB performance improvement over the RIS-SM
system for systems with $N_{r}=4$ and $N_{r}=8$, respectively, at
a BER of $10^{-5}$. It is worth pointing out that the receiver in
the proposed RIS-RQSM system requires minimal CSI due to the use of
the pre-equalizer $G$; this CSI consists only of the average gain
of the effective channel, which is simply a function of the number
of RIS elements, as shown in Section~\ref{sec:System-Model}.

\section{Conclusion\label{sec:Conclusion}}

The RIS-assisted receive quadrature spatial modulation (RIS-RQSM)
system was proposed in this paper as a general approach to RIS-assisted
receive SM with excellent performance. The proposed system increases
the spectral efficiency by implementing both\emph{ }quadrature spatial
modulation \emph{and} IQ modulation, while maintaining the signal
quality at the receiver. In the proposed RIS-RQSM system, the phase
shifts of the RIS elements are designed to construct an IQ symbol
at the receiver; this enables the system to transmit two separate
PAM symbols in the presence of the RIS. We introduced a one-tap pre-equalizer
to allow the proposed low-complexity GD to detect the symbols with
minimum CSI requirement. Analytical results and numerical simulations
both verify the excellent performance of the system and extensively
demonstrate its superiority over comparable benchmark schemes in the
literature. The many advantages of the RIS-RQSM system makes it a
viable candidate for next-generation wireless communication networks.

\appendices{}

\section{Proof of Theorem \ref{thm:ave_Y}\label{sec:proof ave}}

Here we analyze the average of 
\[
Y_{R}^{\star}=\text{sgn}\left(x^{\mathcal{R}}\right)\left(\mathbf{h}_{m}^{\mathcal{R}}\boldsymbol{\theta}^{\mathcal{R}\star}-\mathbf{h}_{m}^{\mathcal{I}}\boldsymbol{\theta}^{\mathcal{I}\star}\right)=\sum_{i=1}^{N}\frac{\lambda A_{i}^{2}+\lambda C_{i}^{2}+\left(1-\lambda\right)A_{i}B_{i}+\left(1-\lambda\right)C_{i}D_{i}}{\sqrt{\left(\lambda A_{i}+\left(1-\lambda\right)B_{i}\right)^{2}+\left(\lambda C_{i}+\left(1-\lambda\right)D_{i}\right)^{2}}}.
\]
As stated before, for large values of $N$, we have $\mathbb{V}\left\{ \lambda\right\} \approx0$;
therefore, we replace $\lambda$ by $\bar{\lambda}$ in calculating
the average of $Y_{R}^{\star}$; this yields {\small{}
\begin{align*}
\mathbb{E}\left\{ Y_{R}^{\star}\right\}  & \approxeq\mathbb{E}\left\{ \sum_{i=1}^{N}\frac{\bar{\lambda}A_{i}^{2}+\bar{\lambda}C_{i}^{2}+\left(1-\bar{\lambda}\right)A_{i}B_{i}+\left(1-\bar{\lambda}\right)C_{i}D_{i}}{\sqrt{\left(\bar{\lambda}A_{i}+\left(1-\bar{\lambda}\right)B_{i}\right)^{2}+\left(\bar{\lambda}C_{i}+\left(1-\bar{\lambda}\right)D_{i}\right)^{2}}}\right\} \\
 & =N\mathbb{E}\left\{ \frac{\bar{\lambda}A_{i}^{2}+\bar{\lambda}C_{i}^{2}+\left(1-\bar{\lambda}\right)A_{i}B_{i}+\left(1-\bar{\lambda}\right)C_{i}D_{i}}{\sqrt{\left(\bar{\lambda}A_{i}+\left(1-\bar{\lambda}\right)B_{i}\right)^{2}+\left(\bar{\lambda}C_{i}+\left(1-\bar{\lambda}\right)D_{i}\right)^{2}}}\right\} ,
\end{align*}
}where we used the fact that each of the summands has an identical
distribution. In the following, we evaluate the average of the terms
in the above summation individually and we omit the index $i$ to
simplify the notation; hence we define 
\[
W_{1}\triangleq\bar{\lambda}\frac{A^{2}}{\sqrt{Z}},\;W_{2}\triangleq\bar{\lambda}\frac{C^{2}}{\sqrt{Z}},\;W_{3}\triangleq\left(1-\bar{\lambda}\right)\frac{AB}{\sqrt{Z}},\;W_{4}\triangleq\left(1-\bar{\lambda}\right)\frac{CD}{\sqrt{Z}},
\]
where $Z\triangleq\left(\bar{\lambda}A+\left(1-\bar{\lambda}\right)B\right)^{2}+\left(\bar{\lambda}C+\left(1-\bar{\lambda}\right)D\right)^{2}$.

According to the law of total expectation, the expected value of $W_{1}$
can be expressed as 
\begin{equation}
\mathbb{E}\left\{ W_{1}\right\} =\mathbb{E}_{A}\left\{ \mathbb{E}_{W_{1}|A}\left\{ W_{1}|A\right\} \right\} =\bar{\lambda}\mathbb{E}_{A}\left\{ A^{2}\mathbb{E}_{Z|A}\left\{ Z^{-\frac{1}{2}}|A\right\} \right\} ,\label{eq:E=00007BW1=00007Ddef}
\end{equation}
where $\mathbb{E}_{Z|A}\left\{ Z^{-\frac{1}{2}}|A\right\} $ is the
inverse-fractional moment of $Z$ where $A$ is given, i.e., where
$A$ is a constant. For a given $A$, using $\bar{\lambda}=\frac{\delta^{2}}{1+\delta^{2}}$
we have $\left(\bar{\lambda}A+\left(1-\bar{\lambda}\right)B\right)\sim\mathcal{N}\left(\bar{\lambda}A,\frac{\bar{\lambda}\left(1-\bar{\lambda}\right)}{2}\right)$,
and $\left(\bar{\lambda}C+\left(1-\bar{\lambda}\right)D\right)\sim\mathcal{N}\left(0,\frac{\bar{\lambda}}{2}\right)$.
Hence, the \ac{RV} $(Z|A)$ is the sum of two independent chi-square
\acp{RV} each having one degree of freedom. The inverse-fractional
moment of $\left(Z|A\right)$ can be computed by using the following
equation \cite{mathai1992quadratic}
\begin{equation}
\mathbb{E}_{Z|A}\left\{ Z^{-c}|A\right\} =\frac{1}{\Gamma\left(c\right)}\int_{0}^{\infty}s^{c-1}\mathbb{E}_{Z|A}\left\{ e^{-sZ}|A\right\} \mathrm{d}s,\label{eq:fractional moment formula}
\end{equation}
where $\mathbb{E}_{Z|A}\left\{ e^{-sZ}|A\right\} =\mathcal{L}_{s}\left(f_{Z}\left(Z|A\right)\right)$
is the \ac{LT} of $f_{Z}\left(Z|A\right)$. We know that the \ac{LT}
of the \ac{PDF} of the sum of independent \acp{RV} is equal to the
product of the LTs of their individual \acp{PDF}, and that the \ac{LT}
of the \ac{PDF} of an \ac{RV} $X=\sum_{i=1}^{n}X_{i}^{2}$ with
$X_{i}\sim\mathcal{N}\left(\mu_{i},\sigma^{2}\right)$ is given by
\begin{equation}
\mathcal{L}_{s}\left(f_{X}(X)\right)=\left(\frac{1}{1+2\sigma^{2}s}\right)^{\frac{n}{2}}\exp\left(\frac{-\mu^{2}s}{1+2\sigma^{2}s}\right),\label{eq:general LT}
\end{equation}
where $\mu^{2}=\sum_{i=1}^{n}\mu_{i}^{2}$. Hence, the \ac{LT} of
$f_{Z}\left(Z|A\right)$ is calculated as 
\begin{equation}
\mathcal{L}_{s}\left(f_{Z}\left(Z|A\right)\right)=\left(\frac{1}{1+\bar{\lambda}s}\right)^{\frac{1}{2}}\left(\frac{1}{1+\bar{\lambda}\left(1-\bar{\lambda}\right)s}\right)^{\frac{1}{2}}\exp\left(\frac{-\bar{\lambda}^{2}A^{2}s}{1+\bar{\lambda}\left(1-\bar{\lambda}\right)s}\right).\label{eq:Laplas(Z|A)}
\end{equation}
Then, (\ref{eq:E=00007BW1=00007Ddef}) can be written as 
\begin{align*}
\mathbb{E}\left\{ W_{1}\right\} = & \frac{\bar{\lambda}}{\Gamma^{2}\left(\frac{1}{2}\right)}\int_{0}^{\infty}s^{\frac{1}{2}-1}\left(\frac{1}{1+\bar{\lambda}s}\right)^{\frac{1}{2}}\left(\frac{1}{1+\bar{\lambda}\left(1-\bar{\lambda}\right)s}\right)^{\frac{1}{2}}\\
 & \times\left(\int_{-\infty}^{\infty}A^{2}\exp\left(-A^{2}\frac{1+\bar{\lambda}s}{1+\bar{\lambda}\left(1-\bar{\lambda}\right)s}\right)\mathrm{d}A\right)\mathrm{d}s,
\end{align*}
where we used the fact that $f_{A}\left(A\right)=\frac{1}{\Gamma\left(\frac{1}{2}\right)}\exp\left(-A^{2}\right)$.
Since $\int_{-\infty}^{\infty}x^{2}\exp\left(-\frac{x^{2}}{2\sigma^{2}}\right)\mathrm{d}x=\Gamma\left(\frac{1}{2}\right)\sigma^{2}\sqrt{2\sigma^{2}}$,
we have 
\[
\int_{-\infty}^{\infty}A^{2}\exp\left(-A^{2}\frac{1+\bar{\lambda}s}{1+\bar{\lambda}\left(1-\bar{\lambda}\right)s}\right)\mathrm{d}A=\frac{\Gamma\left(\frac{1}{2}\right)}{2}\left(\frac{1+\bar{\lambda}\left(1-\bar{\lambda}\right)s}{1+\bar{\lambda}s}\right)^{\frac{3}{2}}.
\]
It follows that 
\begin{align*}
\mathbb{E}\left\{ W_{1}\right\}  & =\bar{\lambda}\frac{1}{2\Gamma\left(\frac{1}{2}\right)}\int_{0}^{\infty}s^{\frac{1}{2}-1}\frac{1+\bar{\lambda}\left(1-\bar{\lambda}\right)s}{\left(1+\bar{\lambda}s\right)^{2}}\mathrm{d}s\\
 & =\bar{\lambda}\left(1-\bar{\lambda}\right)\frac{1}{2\Gamma\left(\frac{1}{2}\right)}\left(\int_{0}^{\infty}s^{\frac{1}{2}-1}\left(1+\bar{\lambda}s\right)^{-1}\mathrm{d}s+\frac{\bar{\lambda}}{1-\bar{\lambda}}\int_{0}^{\infty}s^{\frac{1}{2}-1}\left(1+\bar{\lambda}s\right)^{-2}\mathrm{d}s\right).
\end{align*}
Recalling the definition of the type-2 beta function $\mathrm{B}\left(\alpha,\beta\right)=\int_{0}^{\infty}\frac{t^{\alpha-1}}{\left(1+t\right)^{\alpha+\beta}}\mathrm{d}t=\frac{\Gamma\left(\alpha\right)\Gamma\left(\beta\right)}{\Gamma\left(\alpha+\beta\right)}$,
after some minor manipulations we obtain 
\[
\mathbb{E}\left\{ W_{1}\right\} =\bar{\lambda}^{\frac{1}{2}}\left(1-\bar{\lambda}\right)\frac{1}{2\Gamma\left(\frac{1}{2}\right)}\left(\frac{\Gamma\left(\frac{1}{2}\right)\Gamma\left(\frac{1}{2}\right)}{\Gamma\left(1\right)}+\frac{\bar{\lambda}}{1-\bar{\lambda}}\frac{\Gamma\left(\frac{1}{2}\right)\Gamma\left(\frac{3}{2}\right)}{\Gamma\left(2\right)}\right)=\frac{\sqrt{\pi}}{4}\left(2\bar{\lambda}^{\frac{1}{2}}-\bar{\lambda}^{\frac{3}{2}}\right).
\]
By symmetry it is clear that $\mathbb{E}\left\{ W_{2}\right\} =\mathbb{E}\left\{ W_{1}\right\} $.

Next we determine $\mathbb{E}\left\{ W_{3}\right\} =\mathbb{E}\left\{ \left(1-\bar{\lambda}\right)\frac{AB}{\sqrt{Z}}\right\} $.
Using the law of total expectation, we can write 
\begin{align*}
\mathbb{E}\left\{ W_{3}\right\}  & =\left(1-\bar{\lambda}\right)\mathbb{E}_{A}\left\{ A\mathbb{E}_{B}\left\{ B\mathbb{E}_{Z|(A,B)}\left\{ Z^{-\frac{1}{2}}|(A,B)\right\} \right\} \right\} .
\end{align*}
Given constant $(A,B)$, we have 
\begin{equation}
\mathcal{L}_{s}\left(f_{Z}\left(Z|(A,B)\right)\right)=\left(\frac{1}{1+\bar{\lambda}s}\right)^{\frac{1}{2}}\exp\left(-\left(\bar{\lambda}A+\left(1-\bar{\lambda}\right)B\right)^{2}s\right).\label{eq:Laplace(Z|A,B)}
\end{equation}
Using (\ref{eq:fractional moment formula}), we have 
\[
\mathbb{E}\left\{ W_{3}\right\} =\left(1-\bar{\lambda}\right)\mathbb{E}_{A}\Biggl\{ A\mathbb{E}_{B}\Biggl\{\frac{B}{\Gamma\left(\frac{1}{2}\right)}\int_{0}^{\infty}s^{\frac{1}{2}-1}\left(\frac{1}{1+\bar{\lambda}s}\right)^{\frac{1}{2}}\exp\left(-\left(\bar{\lambda}A+\left(1-\bar{\lambda}\right)B\right)^{2}s\right)ds\Biggr\}\Biggr\}.
\]
Then, using $f_{A}\left(A\right)=\frac{1}{\Gamma\left(\frac{1}{2}\right)}\exp\left(-A^{2}\right)$
and $f_{B}\left(B\right)=\frac{\left(1-\bar{\lambda}\right)^{\frac{1}{2}}}{\bar{\lambda}^{\frac{1}{2}}\Gamma\left(\frac{1}{2}\right)}\exp\left(-\frac{1-\bar{\lambda}}{\bar{\lambda}}B^{2}\right)$,
after some algebraic manipulations we obtain {\small{}
\begin{align}
\mathbb{E}\left\{ W_{3}\right\} = & \frac{\left(1-\bar{\lambda}\right)^{\frac{3}{2}}}{\bar{\lambda}^{\frac{1}{2}}}\frac{1}{\Gamma^{3}\left(\frac{1}{2}\right)}\int_{0}^{\infty}s^{\frac{1}{2}-1}\left(\frac{1}{1+\bar{\lambda}s}\right)^{\frac{1}{2}}\nonumber \\
 & \times\left[\int_{-\infty}^{\infty}A\exp\left(-A^{2}\frac{1+\bar{\lambda}s}{1+\bar{\lambda}\left(1-\bar{\lambda}\right)s}\right)\left(\int_{-\infty}^{\infty}B\exp\left(-\frac{\left(B+\frac{\bar{\lambda}^{2}As}{1+\bar{\lambda}\left(1-\bar{\lambda}\right)s}\right)^{2}}{\frac{\bar{\lambda}/(1-\bar{\lambda})}{\left(1+\bar{\lambda}\left(1-\bar{\lambda}\right)s\right)}}\right)\mathrm{d}B\right)\mathrm{d}A\right]\mathrm{d}s.\label{eq:E=00007BW3=00007D}
\end{align}
}The inner integral over $B$ can be evaluated as {\small{}
\[
\int_{-\infty}^{\infty}B\exp\left(-\frac{\left(B+\frac{\bar{\lambda}^{2}As}{1+\bar{\lambda}\left(1-\bar{\lambda}\right)s}\right)^{2}}{\frac{\frac{\bar{\lambda}}{1-\bar{\lambda}}}{1+\bar{\lambda}\left(1-\bar{\lambda}\right)s}}\right)\mathrm{d}B=-\Gamma\left(\frac{1}{2}\right)\frac{\bar{\lambda}^{\frac{5}{2}}As}{\left(1-\bar{\lambda}\right)^{\frac{1}{2}}\left(1+\bar{\lambda}\left(1-\bar{\lambda}\right)s\right)^{\frac{3}{2}}}.
\]
}Substituting this into (\ref{eq:E=00007BW3=00007D}), the average
of $W_{3}$ is given by 
\begin{align*}
\mathbb{E}\left\{ W_{3}\right\}  & =\bar{\lambda}^{2}\left(1-\bar{\lambda}\right)\frac{-1}{2\Gamma\left(\frac{1}{2}\right)}\int_{0}^{\infty}s^{\frac{3}{2}-1}\frac{1}{\left(1+\bar{\lambda}s\right)^{2}}\mathrm{d}s\\
 & =\bar{\lambda}^{\frac{1}{2}}\left(1-\bar{\lambda}\right)\frac{-1}{2\Gamma\left(\frac{1}{2}\right)}\mathrm{B}\left(\frac{3}{2},\frac{1}{2}\right)=-\frac{\sqrt{\pi}}{4}\bar{\lambda}^{\frac{1}{2}}\left(1-\bar{\lambda}\right).
\end{align*}
Also, by symmetry we have $\mathbb{E}\{W_{4}\}=\mathbb{E}\{W_{3}\}$.
Finally, the average of $Y_{R}^{\star}$ is given by 
\[
\mathbb{E}\left\{ Y_{R}^{\star}\right\} \approx2N\left(\mathbb{E}\left\{ W_{1}\right\} +\mathbb{E}\left\{ W_{3}\right\} \right)=\bar{\lambda}^{\frac{1}{2}}\frac{N\sqrt{\pi}}{2}.
\]
Then, using $\bar{\lambda}=\frac{\delta^{2}}{1+\delta^{2}}$, $\mathbb{E}\left\{ Y_{I}^{\star}\right\} $
is given by 
\[
\mathbb{E}\left\{ Y_{I}^{\star}\right\} =\frac{1}{\delta}\mathbb{E}\left\{ Y_{R}^{\star}\right\} \approx(1-\bar{\lambda})^{\frac{1}{2}}\frac{N\sqrt{\pi}}{2}.
\]

\bibliographystyle{ieeetr}
\bibliography{ref}

\end{document}

%% file: acro.tex
\acrodef{5G}{fifth-generation}
\acrodef{6G}{sixth-generation}
\acrodef{ABEP}{average bit error probability}
\acrodef{AO}{alternating optimization}
\acrodef{APGM}{accelerated projected gradient method}
\acrodef{ASP}{antenna separation product}
\acrodef{AWGN}{additive white Gaussian noise}
\acrodef{BEP}{bit error probability}
\acrodef{BER}{bit error rate}
\acrodef{BF-MIMO}[BF\mbox{-}MIMO]{beamforming MIMO}
\acrodef{BF}{beamforming}
\acrodef{bpcu}{bits per channel use}
\acrodef{CDF}{cumulative distribution function}
\acrodef{CF}{characteristic function}
\acrodef{CLT}{central limit theorem}
\acrodef{CP}{cyclic prefix}
\acrodef{CSI}{channel state information}
\acrodef{CSIR}{channel state information at receiver}
\acrodef{CSIT}{channel state information at the transmitter}
\acrodef{DCMC}{discrete\mbox{-}input continuous\mbox{-}output memoryless channel}
\acrodef{DFT}{discrete Fourier transform}
\acrodef{DL-TR-GSM}{dual-layered transmit-receive \ac{GSM}}
\acrodef{DLT}{dual-layered transmission}
\acrodef{EGC}{equal gain combining}
\acrodef{EM}{electromagnetic}
\acrodef{FSPL}{free space path loss}
\acrodef{FFT}{fast Fourier transform}
\acrodef{FDE}{frequency domain equalization}
\acrodef{GD}{greedy detector}
\acrodef{GRSM}{generalized \acl{RSM}}
\acrodef{GSM}{generalized \acl{SM}}
\acrodef{IFFT}{invserse fast Fourier transform}
\acrodef{ICI}{inter-channel interference}
\acrodef{iid}[i.i.d.]{independent and identically distributed}
\acrodef{IM}{index modulation}
\acrodef{IQ}{in\mbox{-}phase and quadrature}
\acrodef{ISI}{intersymbol interference}
\acrodef{ISI-free}[ISI\mbox{-}free]{intersymbol interference free}
\acrodef{KKT}{Karush\mbox{-}Kuhn\mbox{-}Tucker}
\acrodef{LDS}{low density signature}
\acrodef{LIS}{large intelligent surface}
\acrodef{LOS}{line\mbox{-}of\mbox{-}sight}
\acrodef{LT}{Laplace transform}
\acrodef{mmWave}{millimeter-wave}
\acrodef{mMIMO}{massive \ac{MIMO}}
\acrodef{MGF}{moment generating function}
\acrodef{MIMO}{multiple\mbox{-}input multiple\mbox{-}output}
\acrodef{MISO}{multiple\mbox{-}input single\mbox{-}output}
\acrodef{ML}{maximum likelihood}
\acrodef{MP}{message\mbox{-}passing}
\acrodef{MRC}{maximal ratio combining}
\acrodef{MMSE}{minimum mean square error}
\acrodef{MSTBC}{multi\mbox{-}dimensional \ac{STBC}}
\acrodef{MU-TR-GSM}{multiuser transmit-receive  \ac{GSM} }
\acrodef{NCSIT}{no channel state information at TX}
\acrodef{NLOS}{non\mbox{-}\acs{LOS}} 
\acrodef{NOMA}{non-orthogonal multiple access}
\acrodef{OFDM}{orthogonal frequency division multiplexing}
\acrodef{OFDMA}{orthogonal frequency division multiple access}
\acrodef{PA}{power amplifier}
\acrodef{PAE}{power added efficiency}
\acrodef{PAPR}{peak\mbox{-}to\mbox{-}average power ratio}
\acrodef{PDF}{probability density function}
\acrodef{PEP}{pairwise error probability}
\acrodefplural{PEP}{pairwise error probabilities}
\acrodef{PGM}{projected gradient method}
\acrodef{PMP}{probability mass function}
\acrodef{PSM}{precoding-aided spatial modulation}
\acrodef{QSM}{quadrature spatial modulation}
\acrodef{QSSK}{quadrature space-shift keying}
\acrodef{RC}{reorganization computation}
\acrodef{RE}{resource element}
\acrodef{RF}{radio frequency}
\acrodef{RIS}{reconfigurable intelligent surface}
\acrodef{RIS-AP}{RIS-access point}
\acrodef{RIS-RQRM}{RIS\mbox{-}aided receive quadrature reflecting modulation}
\acrodef{RIS-RQSSK}[\ac{RIS}\mbox{-}RQSSK]{\ac{RIS}\mbox{-}assisted receive quadrature space-shift keying}
\acrodef{RIS-SM}{\ac{RIS}-\acl{SM}}
\acrodef{RIS-SSK}{\ac{RIS}\mbox{-}\acl{SSK}}
\acrodef{RPM}{reflection pattern modulation}
\acrodef{RSM}{receive spatial modulation}
\acrodef{RV}{random variable}
\acrodef{RX}{receiver}
\acrodef{SCMA}{sparse code multiple access}
\acrodef{SEP}{symbol error probability}
\acrodef{SER}{symbol error rate}
\acrodef{SINR}{signal-to-interference-plus-noise ratio}
\acrodef{SISO}{single-input single-output}
\acrodef{SL}{sparse layering}
\acrodef{SL-MIMO}{sparse layered \ac{MIMO}}
\acrodef{SM}{spatial modulation}
\acrodef{SMX-MIMO}[SMX\mbox{-}MIMO]{spatial multiplexing MIMO}
\acrodef{SMX}{spatial multiplexing}
\acrodef{SNR}{signal-to-noise ratio}
\acrodef{SC}{single carrier}
\acrodef{SVD}{singular value decomposition}
\acrodef{SPST}{single pole single-throw}
\acrodef{SSK}{space-shift keying}
\acrodef{STBC}{space\mbox{-}time block code}
\acrodef{SU}{secondary user}
\acrodef{TDE}{time domain equalization}
\acrodef{TX}{transmitter}
\acrodef{ULA}{uniform linear array}
\acrodef{URA}{uniform rectangular array}
\acrodef{VBLAST}{vertical Bell Labs layered space\mbox{-}time}
\acrodef{VGA}{variable gain amplifier}
\acrodef{ZF}{zero-forcing}
\acrodef{ZMCG}{zero-mean complex Gaussian}